\documentclass[iop, apj]{emulateapj}

\usepackage{amsmath}
\usepackage{amsfonts}
\usepackage{amssymb}
\usepackage{float}
\usepackage{graphicx}
\usepackage{color, soul}

\newcommand\eps{{\epsilon}}

\newcommand{\grad}{\boldsymbol{\nabla}}

\newcommand{\del}{\partial}
\newcommand{\gdet}{\sqrt{-g}}

\newcommand{\beq}{\begin{equation}}
\newcommand{\eeq}{\end{equation}}
\newcommand{\beqn}{\begin{eqnarray}}
\newcommand{\eeqn}{\end{eqnarray}}

\def\bhat{\boldsymbol{\hat{b}}}

\def\bh{\hat{b}}
\def\bv{\mathbf v}
\def\bq{\mathbf q}

\slugcomment{Accepted for publication in ApJ}

\shortauthors{Chandra et al.}
\shorttitle{Extended MHD}

\begin{document}

\title{An Extended Magnetohydrodynamics Model for Relativistic { Weakly Collisional} Plasmas}

\author{Mani Chandra}
\affil{Department of Astronomy, University of Illinois, 1110 West Green Street, Urbana, IL, 61801}
\email{manic@illinois.edu}
\author{Charles F. Gammie}
\affil{Department of Astronomy, University of Illinois, 1110 West Green Street, Urbana, IL, 61801}
\affil{Department of Physics, University of Illinois, 1110 West Green Street, Urbana, IL, 61801}
\email{gammie@illinois.edu}
\author{Francois Foucart}
\affil{Lawrence Berkeley National Laboratory, 1 Cyclotron Rd, Berkeley, CA 94720, USA; Einstein Fellow}
\email{fvfoucart@lbl.gov}
\author{Eliot Quataert}
\affil{Department of Astronomy and Theoretical Astrophysics Center, University of California, Berkeley, CA, 9472}
\email{eliot@berkeley.edu}

\begin{abstract}

Black holes that accrete far below the Eddington limit are believed to accrete through a geometrically thick, optically thin, rotationally supported plasma that we will refer to as a radiatively inefficient accretion flow (RIAF).  RIAFs are typically collisionless in the sense that the Coulomb mean free path is large compared to $GM/c^2$, and relativistically hot near the event horizon.  In this paper we develop a phenomenological model for the plasma in RIAFs, motivated by the application to sources such as Sgr A* and M87. { The model is derived using Israel-Stewart theory, which considers deviations up to second order from thermal equilibrium, but modified for a magnetized plasma. This leads to thermal conduction along magnetic field lines and a difference in pressure, parallel and perpendicular to the field lines (which is equivalent to anisotrotropic viscosity).} In the non-relativistic limit, our model reduces to the widely used Braginskii theory of magnetized, { weakly collisional} plasmas. We compare our model to the existing literature on dissipative relativistic fluids, describe the linear theory of the plasma, and elucidate the physical meaning of the free parameters in the model.  We also describe limits of the model when the conduction is saturated and when the viscosity implies a large pressure anisotropy.  In future work, the formalism developed in this paper will be used in numerical models of RIAFs to assess the importance of non-ideal processes for the dynamics and radiative properties of slowly accreting black holes.
\end{abstract}

\section{Introduction and Astrophysical Context}

Most massive galaxies have black holes at their centers, and most of these black holes are accreting far below the Eddington rate $\dot{M}_{Edd}$ (\citealt{Ho2009}).  Low luminosity black holes are believed to accrete through a geometrically thick, optically thin disk.  Phenomenological (radiatively inefficient accretion flows or RIAFs, see \citealt{Yuan2014}) and numerical (general relativistic magnetohydrodynamics, or GRMHD, see e.g. \citealt{Koide1999,DeVilliers2003,McKinney2004}) models suggest that the density and temperature of the accreting plasma are such that the collisional (Coulomb) mean free paths (ion-ion, ion-electron, and electron-electron) are many orders of magnitude larger than $GM/c^2$ when $\dot M \ll \dot M_{Edd}$ \citep{Mahadevan1997}.  The accreting plasma is thus {\em collisionless}.

In the nearby universe, the roster of low luminosity black holes includes M87 (accreting 4-5 orders of magnitude below $\dot{M}_{Edd}$), and Sgr A* (accreting about 8 orders of magnitude below $\dot{M}_{Edd}$).  These two sources are the largest known black holes in terms of angular size on the sky.  As a result, they are the two main targets for high resolution imaging experiments, including Gravity on the VLT \citep{Eisenhauer2008} and the Event Horizon Telescope (EHT; \citealt{Doeleman2009}); the latter  will use submillimeter very long baseline interferometry (VLBI) to resolve the accretion flow and jet on angular scales comparable to the event horizon.  The EHT data may be used to test General Relativity by measuring the angular size of the photon orbit, if astrophysical uncertainties can be controlled
\citep{Psaltis2014}.

{ The fact that the collisional mean free path is much larger than $GM/c^2$ in RIAFs implies that non-ideal processes such as conduction and viscosity are likely to be important. Furthermore, the mean free path is also much larger than the Larmor radii of all the species in the plasma and the gyration time scale is much shorter than the dynamical time scale. This leads to the above dissipative processes being anisotropic with respect to the local magnetic field. Heat flows only along the field lines and an anisotropic viscosity is generated by a shear flow projected along the field lines.}

Consider conduction, using non-relativistic estimates: the volume heating rate is $du/dt = -\grad \cdot \bq$, where $\bq$ is the conductive heat flux. The timescale for changing the internal energy is $\tau_{cond} = u/(du/dt)$.  Estimate $\grad \cdot \bq \sim q/r$, assume the heat flux is approximately saturated (given the large mean free paths) so that $q \sim u v$, where $u$ is the internal energy density and $v$ is rms particle speed.  Then  $\tau_{cond} \sim r/c$ for electrons (electrons are relativistic close to the horizon, so $v \sim c$) and  $\tau_{cond} \sim r/c_s$ for protons, which in a near-virial RIAF is the dynamical time.  Conduction can thus potentially play an important role in controlling the thermal state of the accreting plasma (e.g., \citealt{Johnson2007}).

{ What about viscosity or equivalently (as we show below) a difference in pressures along and perpendicular to the local magnetic field? Pressure anisotropy can be generated in a number of ways, for example, by anisotropic compression or expansion of the plasma, through a linear shear, etc.} If the magnetic field strength varies, adiabatic invariance of the magnetic moment associated with the orbit of charged particles about a magnetic field with strength $B$ implies that $T_\perp/B$ is invariant for a plasma with $kT \lesssim m c^2$ ($T_\perp \equiv$  temperature perpendicular to the local magnetic field). Here, the temperature $T$ and the mass $m$ corresponds to a specific species. Thus, if the plasma is compressed in the plane perpendicular to the mean field so that the density increases by a factor $R$, the perpendicular temperature also increases by a factor $R$, generating a significant pressure anisotropy, i.e., viscosity.  Order unity fluctuations in density and magnetic field strength are common in numerical models of accretion flows, so order unity pressure anisotropy is expected in the absence of collective effects.  This implies that viscous stresses may be dynamically important and contribute significantly to angular momentum transport and plasma heating (\citealt{Sharma2006, Sharma2007}).

Despite their potential importance, conduction and viscosity are, however, absent from all global relativistic numerical models of RIAFs to date.   This is one of the significant systematic uncertainties in developing models for the emission from systems such as Sgr A* and M87.  

In this paper, we develop a formalism for modeling relativistic anisotropic conduction and viscosity, motivated by the application to RIAFs.   Although the plasmas of interest are macroscopically collisionless, we focus in this paper on the more modest task of developing a theory for collisional magnetized plasmas in which the mean free path is large compared to the Larmor radius of particles, but small compared to the system size.  The former hierarchy implies that heat and momentum transport are predominantly along the local magnetic field direction, while the latter constraint implies that one can derive the relevant equations using an expansion about thermal equilibrium.  Our assumed hierarchy of length-scales is similar to that used in \citealt{Braginskii1965}'s theory of non-relativistic magnetized plasmas, which has been widely applied to understand the physics of dilute astrophysical plasmas (see \citealt{Kulsrud2005}).  We will refer to our formalism as an extended MHD (EMHD) model.

Although the applications that motivate this work are to collisionless systems, we focus on the collisional regime for the following reasons:  (1) the theory of dissipative relativistic fluids is quite subtle (e.g. \citealt{Andersson2007}), so it seems prudent to not jump directly to the yet more challenging long mean free path regime;  (2) wave-particle interactions and velocity space instabilities  limit the mean free path of charged particles to be much less than the collisional mean free path under the conditions of interest (\citealt{Sharma2006,Kunz2014, Riquelme2015}) implying that the `collisional' theory may be more appropriate than one might have first anticipated.   The formalism we develop allows  the viscosity and conductivity to depend arbitrarily on local plasma conditions so that these wave-particle limits on the mean free path can be incorporated as sub-grid models.

Throughout the paper we formulate the equations in terms of a single fluid model.  In reality, the low-collisionality plasmas of interest are believed to develop a two-temperature structure because the timescale for Coulomb collisions to equilibrate the electron and proton temperatures is long compared to the dynamical time in the accreting plasma. A formulation for dealing with electron dynamics and its numerical implementation has been recently introduced by \cite{Ressler2015}, where a reduced form of our conduction model has been used and appropriately modified for electrons.

The remainder of this paper is organized as follows.  In \S 2 we write down basic equations and describe the equivalence of viscosity and anisotropic pressure.  In \S 3 we describe the desired asymptotic behavior of any extended MHD closure model.  In \S 4 we derive the evolution equations for the heat flux and pressure anisotropy. { \S 5 describes the connection between our model and non-relativistic dissipative theory.}  \S 6 motivates a scheme for fixing the model parameters (the transport coefficients : viscosity and conductivity) in terms of a relaxation time.  \S 7 gives the linear theory and the stability thresholds of our model, and \S 8 a brief discussion of nonlinear (shock) solutions.  Finally in \S 9 we offer a guide to the model, a summary of the formalism, and the relationship to earlier works.  In the Appendix we show how the characteristic pressure anisotropy derived on thermodynamic grounds using the Israel-Stewart theory can also be interpreted as arising from conservation of relativistic adiabatic invariants in a magnetized plasma.

\section{Physical Context}

We begin by defining notation and frames.  

{ We work in a spacetime described by the metric $g_{\mu \nu}$, whose determinant we denote with $g$. Consider a plasma consisting of particles with distribution function $f_s \equiv dN/d^3x d^3p$, and rest mass $m_s$, where $d^3p = dp_1 dp_2 dp_3$, $p_i$ are the spatial {\em covariant} components of the particle four-momentum $p_\mu$ and $s$ indicates the species of the particles (electrons, ions, etc.).  The distribution function is invariant. Each species has a number current
\begin{equation}
N^\mu_s \equiv \int \, \frac{d^3 p}{\gdet p^t} \, p^\mu f_s.
\end{equation}
We assume that the plasma consists of electrons and ions, and is quasi-neutral everywhere. Thus both these species have the same approximate number current $N^\mu \equiv N^\mu_i \approx N^\mu_e$. We define the rest frame as that in which the number current has no spatial components. Therefore we have,
\begin{equation} \label{eq:mattercurrent}
u^\mu \equiv \frac{N^\mu}{n}.
\end{equation}
where $n$ is the number density of ions, and is equal to the number density of electrons. This definition of $u^\mu$ implies that we are using the so-called Eckart frame, in which mass diffusion is absent.  An alternative, the Landau frame, assumes that energy diffusion is absent. The total rest mass density $\rho$, often denoted as $\rho_0$ in the relativity literature is $\rho = -\Sigma_s m_s N^\mu_s u_\mu = -(m_i N^\mu_i u_\mu + m_e N^\mu_e u_\mu)$. Since $N^\mu_i \approx N^\mu_e \equiv N^\mu$, we have $\rho = -(m_i + m_e) N^\mu u_\mu = (m_i + m_e)n$.
}

The matter stress-energy tensor is
\begin{equation}\label{eq:stressenergy}
T^{\mu\nu}_{\mathrm{matter}} \equiv \sum_s \int \, \frac{d^3 p}{\gdet p^t} \, p^\mu p^\nu f_s.
\end{equation}
Each of these definitions is invariant because $d^3p/(\gdet p^t)$ is invariant. 

On taking moments of the { Boltzmann} equation, one can show that the quantities { $N^\mu$ and $T^{\mu\nu}_{\mathrm{matter}}$} satisfy
{
\begin{equation}
\nabla_\mu N^\mu = \nabla_\mu (n u^\mu) = 0
\label{eq:continuity}
\end{equation}
}
and 
\begin{equation}
\nabla_\mu T^{\mu\nu}_{\mathrm{matter}} = F^{\mu \nu} J_\mu \label{eq:divMatter}
\end{equation}
where $\nabla_\mu$ is the covariant derivative, $F^{\mu \nu}$ is the electromagnetic field tensor and $J_\mu$ is the four-current. The divergence of the electromagnetic stress tensor is 
\beq
\nabla_\mu T^{\mu \nu}_{\mathrm{EM}} = -F^{\mu \nu} J_\mu \label{eq:divEM}
\eeq
On adding (\ref{eq:divMatter}) and (\ref{eq:divEM}), we get that the divergence of the total stress tensor $T^{\mu \nu}$ is zero, as required by the Bianchi identities
\beq
\nabla_\mu \left(T^{\mu \nu}_{\mathrm{matter}} + T^{\mu \nu}_{\mathrm{EM}} \right) =
\nabla_\mu T^{\mu \nu} = 0
\label{eq:divT}
\eeq


For a perfect unmagnetized fluid the stress-energy tensor is
\beq
T^{\mu\nu}_{\mathrm{matter}} = (\rho + u)u^\mu u^\nu + P h^{\mu\nu},
\eeq
where
\beq
h^{\mu\nu} \equiv g^{\mu\nu} + u^\mu u^\nu
\eeq
is the projection tensor (projects into the space normal to the fluid four-velocity), $g^{\mu\nu}$ is the metric, $u$ the internal energy per unit proper volume, and $P$ the gas pressure. 

We define the {\em fluid frame} as an orthonormal tetrad with time component $e_{(t)}^\mu = u^\mu$, and three additional spacelike basis vectors $e_{(x)}^\mu, e_{(y)}^\mu$, and $e_{(z)}^\mu$.
In the fluid frame the stress-energy tensor is
\begin{equation}
T^{(a)(b)}_{\mathrm{matter}} = \left (
\begin{array}{cccc}
\rho + u & 0 & 0 & 0 \\
0 & P & 0 & 0 \\
0 & 0 & P & 0 \\
0 & 0 & 0 & P 
\end{array}
\right),
\end{equation}
which, with (\ref{eq:stressenergy}), provides a kinetic theory definition for the pressure and internal energy.

The space-space part of the stress-energy tensor is the pressure, or stress, tensor
\begin{equation}\label{eq:ptensdef}
P^{\mu\nu} \equiv h^\mu_\alpha h^\nu_\beta T^{\alpha\beta}.
\end{equation}
In the fluid frame, the spatial components of the ideal fluid pressure tensor are
\begin{equation}
P^{ij} = \left (
\begin{array}{ccc}
P & 0 & 0 \\
0 & P & 0 \\
0 & 0 & P 
\end{array}
\right).
\end{equation}
In what follows we are interested in modeling a magnetized plasma that departs from ideality in that it has a conductive heat flux and a viscous stress. The total stress tensor $T^{\mu \nu}$ with the fluid assumed perfect and the electromagnetic terms included under the ideal magnetohydrodynamics (MHD) approximation (conductivity $\sigma = \infty$) is
\beq
T^{\mu\nu} = (\rho + u + \frac{1}{2} b^2)u^\mu u^\nu + (P + \frac{b^2}{2}) h^{\mu\nu} - b^\mu b^\nu.
\eeq
where $b^2 = b^\mu b_\mu$ and 
\begin{equation}
b^\mu = \frac{1}{2} \eps^{\mu\nu\kappa\lambda} u_\nu F_{\lambda\kappa}
\end{equation}
where $\eps \equiv$ Levi-Civita tensor, which is antisymmetric on all pairs of indices.  Evidently $b^\mu u_\mu = 0$, and $b^\mu$ reduces to the magnetic field in the fluid frame (with a factor of $\sqrt{4\pi}$ absorbed into the definition).  The magnetic field evolution is given by
\begin{equation}
\nabla_\mu \left(u^\mu b^\nu - b^\mu u^\nu\right) = 0
\label{eq:induction}
\end{equation}
which combines the induction equation (three space components) with the no-monopoles condition (time component).

When including non-ideal effects in the stress-energy tensor, the heat flux $q^\mu$ in the system is
\begin{equation}\label{eq:qflux}
q^\mu \equiv - h^\mu_\alpha u_\beta T^{\alpha\beta}.
\end{equation}
Combined with (\ref{eq:stressenergy}), this provides a kinetic theory definition for the heat flux.  Evidently $u^\mu q_\mu = 0$, so in the fluid frame, $q^{(t)} = 0$.  From (\ref{eq:qflux}) one can show that the heat flux makes a contribution to the stress tensor of the form 
\begin{equation}\label{eq:condSE}
T^{\mu\nu}_{\mathrm{cond}} = u^\mu q^\nu + u^\nu q^\mu,
\end{equation}
which in the fluid frame has the form
\begin{equation}
T^{(a)(b)}_{\mathrm{cond}} = \left (
\begin{array}{cccc}
0 & q^x & q^y & q^z \\
q^x & 0 & 0 & 0 \\
q^y & 0 & 0 & 0 \\
q^z & 0 & 0 & 0
\end{array}
\right).
\end{equation}

The viscous stress tensor $\Pi^{\mu \nu}$ models momentum fluxes set up by departures from equilibrium due to a shear flow.  It is given by
\begin{equation}
\Pi^{\mu\nu} + P h^{\mu\nu} \equiv h^\mu_\alpha h^\nu_\beta T^{\alpha\beta},
\end{equation}

It is perhaps not as widely appreciated as it should be, that the viscous stress can be recast as a pressure anisotropy.  In the fluid frame the viscous stress tensor $\tau^{ij}$ is a symmetric matrix, so there is always a basis (obtained by rotation) where it can be written in diagonal form:
\begin{equation}
P^{ij} = \left (
\begin{array}{ccc}
P_x & 0 & 0 \\
0 & P_y & 0 \\
0 & 0 & P_z 
\end{array}
\right),
\end{equation}
where there is a separate pressure for each direction.  Below we consider a magnetized plasma where the stress tensor is symmetric under rotations around the magnetic field.  If the field is in the $z$ direction, this implies $P_x = P_y$.  

With conduction and viscosity of the plasma included, the total stress-energy tensor is now
\begin{widetext}
\beq
T^{\mu \nu}_{\mathrm{matter+EM}} = T^{\mu\nu} = (\rho + u + \frac{1}{2} b^2)u^\mu u^\nu + (P + \frac{1}{2} b^2) h^{\mu\nu} - b^\mu b^\nu + q^\mu u^\nu + q^\nu u^\mu + \Pi^{\mu\nu}.
\label{eq:fullSEtens}
\eeq
\end{widetext}
This stress-energy tensor is quite general, but one needs an appropriate model for $u$, $P$, $q^\mu$, and $\tau^{\mu\nu}$. Note that the electromagnetic terms are still written down in the infinite conductivity limit.

\section{Model Desiderata}

What are the desirable properties of a closure model for the heat flux $q^\mu$ and the viscous stress $\Pi^{\mu\nu}$?

(1) The model should be causal.
For non-relativistic shear viscosity and thermal conduction, the energy and momentum fluxes are proportional to gradients of the temperature and velocity, and so respond instantaneously to changes in the fluid.  Classical, non-relativistic models are parabolic and have characteristics that propagate at infinite speed.  The classical theories can be made causal in a model pioneered by Maxwell and \cite{Cattaneo1948} in which the energy and momentum fluxes relax to their classical values on a characteristic timescale $\tau$.  

(2) The model should be stable.
The relativistic thermal conduction model of \cite{Eckart1940} (see also MTW) sets
{\begin{equation}
q^\mu =  -\rho \chi \, h^{\mu\nu} \left( \del_\nu \Theta + \Theta a_\nu\right)
\end{equation}
 where $\chi$ is the thermal diffusivity, $\Theta = kT/mc^2$ is the normalized temperature and $a_\nu$ is the four-acceleration.}  The term proportional to the four-acceleration drives the temperature toward a constant redshifted temperature rather than a constant local temperature - a desirable effect - but it makes the theory unstable (\citealt{Garcia2009, Lopez2011}).  Long wavelength modes ($k\rightarrow0$) are unstable with growth rate
\begin{eqnarray}
\omega = \frac{(\rho c^2 + u + P)c^2}{\chi P}
\end{eqnarray}
Notice that as $\chi \rightarrow 0$, $\omega \rightarrow \infty$.  If this theory were correct, the water in our bodies would spontaneously explode in $10^{-34}$ sec (\citealt{Hiscock1985}).  Evidently the stability of relativistic conduction theories is nontrivial.  A relativistic extension of the Maxwell-Cattaneo procedure not only makes the theory hyperbolic but also conditionally eliminates the Eckart instability.

(3) Entropy should increase, i.e. the model should obey the second law of thermodynamics.
The entropy constraint is expressed by defining an entropy four-current $s^\mu$ and requiring that $s^\mu_{;\mu} \ge 0$.  This constraint was used to derive the Eckart model wherein the entropy current is expanded around equilibrium to first order in the heat flux. The first order model suffers from the instability described above. Expanding up to second order, as done by \cite{Israel1979} leads to { conditionally hyperbolic,} stable and causal equations. We will use this in the next section to derive evolution equations for our model of anisotropic thermal conduction and viscosity.

(4) We are interested in plasmas with ion and electron Larmor radii tiny compared to the characteristic scale $GM/c^2$, and ion and electron gyroperiod tiny compared to the dynamical timescale. Therefore, we shall assume that the distribution functions of both the ions and electrons are independent of the gyrophase, i.e. $f_s \equiv f_s(p_\parallel, p_\perp)$, where $s$ indicates the species. Now in a tetrad frame with $e^\mu_{(t)} = u^\mu$ and $e^\mu_{(z)} \equiv e^\mu_{\parallel} = \hat{b}^\mu$, we evaluate (\ref{eq:stressenergy}) to find only the following non-zero components
\begin{equation}
T^{(\mu)(\nu)}_{\mathrm{matter}} = \left (
\begin{array}{cccc}
T^{(0)(0)} & 0 & 0 & T^{0\parallel} \\
0 & T^{\perp \perp} & 0 & 0 \\
0 & 0 & T^{\perp \perp} & 0 \\
T^{0 \parallel} & 0 & 0 & T^{\parallel \parallel}
\end{array}
\right).
\end{equation}
The terms which are zero are identically so, because they appear as $\int_0^{2\pi} \sin(\theta) d \theta$ or $\int_0^{2\pi} \cos(\theta) d \theta$, where $\theta$ is the gyrophase. We see that there is a heat flux $T^{0\parallel} \equiv q$ only along the magnetic field line. Therefore our model for the heat flux can be written as
\begin{equation}
q^\mu = q\, \hat{b}^\mu \label{eq:qmodel}
\end{equation}
where $\hat{b}^\mu = b^\mu/\sqrt{b^\mu b_\mu}$.  $q$ will be the fundamental variable describing the heat flux.\footnote{ { A more accurate description of a collisionless plasma requires us to differentiate between a heat flow due to parallel temperatures gradients $q^\mu_\parallel$ and a heat flow due to perpendicular temperatures gradients $q^\mu_\perp$, both of which flow along the field lines $q^\mu_\parallel \equiv q_\parallel\,\hat{b}^\mu$ and $q^\mu_\perp \equiv q_\perp\,\hat{b}^\mu$. The net heat flow is then $q^\mu = q\,\hat{b}^\mu \equiv (q_\perp + q_\parallel)\,\hat{b}^\mu$. However, even in this case, the heat flux appears in the stress tensor only as the sum $q \equiv q_\perp + q_\parallel$.} }

We now write down the pressure tensor with $T_{\perp \perp} \equiv P_\perp$ and $T_{\parallel \parallel} \equiv P_\parallel$
\begin{equation}
P^{(i)(j)} = 
\left (
\begin{array}{ccc}
P_\perp & 0 & 0 \\
0 & P_\perp & 0 \\
0 & 0 & P_\parallel 
\end{array}
\right) =
\left (
\begin{array}{ccc}
P + \Delta P_{\perp} & 0 & 0 \\
0 & P + \Delta P_{\perp} & 0 \\
0 & 0 & P + \Delta P_{\parallel}
\end{array}
\right).
\end{equation}
where $P$ is the ideal fluid pressure and $\Delta P_{\perp}$ and $\Delta P_\parallel$ are deviations from it in the $\perp$ and $\parallel$ directions respectively. The variables $\Delta P_{\perp}$ and $\Delta P_\parallel$ can in principle vary independently and give rise to both a bulk viscosity (trace) and a shear viscosity (trace-free part). We simplify the model further by assuming that bulk viscosity is zero and thus imposing that the deviation of the pressure tensor from ideality be trace-free. Doing so gives $\Delta P_\parallel = -2 \Delta P_\perp$. Now redefining $\Delta P_\perp \equiv \Delta P/3$, we have for the pressure tensor
\begin{equation}
P^{(i)(j)} = 
\left (
\begin{array}{ccc}
P + \frac{1}{3}\Delta P & 0 & 0 \\
0 & P + \frac{1}{3} \Delta P & 0 \\
0 & 0 & P - \frac{2}{3}\Delta P 
\end{array}
\right).
\end{equation}
We see that $P_\perp = P + \Delta P/3$ and $P_\parallel = P - 2\Delta P/3$. Therefore $\Delta P = P_\perp - P_\parallel$, which is the usual definition of pressure anisotropy.
The shear stress in an arbitrary frame is then
\begin{equation}
\Pi^{\mu\nu} = -\Delta P \left(\hat{b}^\mu \hat b^{\nu} - \frac{1}{3} h^{\mu\nu}\right).
\label{eq:taumodel}
\end{equation}
$\Delta P$ is the fundamental variable describing the viscous stress. The above expression satisfies $\Pi^{\mu}_\mu=0$ and is thus trace-free.

(5) If possible, the model should asymptote to a rigorous model in the collisional limit.
The relation to earlier theories will be discussed in detail later, but in brief our model is equivalent to \cite{Israel1979} theory projected along the magnetic field lines.  Israel-Stewart theory has 9 fields that describe nonideal effects: 5 for the shear viscosity, 1 for bulk viscosity, and 3 for the conductivity.  Projecting along the magnetic field lines reduces the viscous shear stress degrees of freedom from 5 to 1, and the heat flux degrees of freedom from 3 to 1, while we ignore bulk viscosity.  

\section{Evolution of $q$ and $\Delta P$}
\label{sec:evolution}

Following Israel and Stewart, it is possible to derive evolution equations for $q$ and $\Delta P$ from the second law of thermodynamics, expressed here by the requirement that the entropy current $s^\mu$ have positive divergence: $\nabla_\mu s^\mu \ge 0$.  

First, what is the entropy current?  In ideal hydrodynamics
\begin{equation}
s^\mu = s \rho u^\mu
\end{equation}
where the {entropy per baryon} $s$ depends on the equation of state $P = P(\rho,u)$.  Here we assume
\begin{equation}
P = (\gamma - 1) u
\end{equation}
and
\begin{equation}
P = \rho \Theta.
\end{equation}
The first law then implies $ds = (du/u - \gamma  d \rho / \rho)/(\gamma-1)$.  One can show that, if $\tau\equiv$ proper time, 
\begin{equation}
0 = u_\nu \nabla_\mu T_{\mathrm{ideal}}^{\mu\nu} = -P \frac{ds}{d\tau}
\label{eq:dsdtau}
\end{equation}
where $T_{\mathrm{ideal}}^{\mu\nu}$ is the ideal gas stress-energy tensor.  Combining this result with the continuity equation gives
\begin{equation}
\nabla_\mu s^\mu = 0.
\end{equation}

In nonideal hydrodynamics one thinks of { $q^\mu$ and $\Pi^{\mu \nu}$} as small corrections to the ideal model. Expanding to second order in these small corrections, the most general possible entropy current { subject to the constraints $q_\mu u^\mu = 0$, $\Pi^\mu_\nu u^\nu=0$ and $\Pi^\mu_\mu=0$} is:
\begin{equation}
s^\mu = s \rho u^\mu 
+ \frac{a_1}{\Theta} q^\mu 
- \frac{b_1}{2\Theta} q^\alpha q_\alpha u^\mu
- \frac{b_2}{2\Theta} \Pi^{\alpha\beta} \Pi_{\alpha\beta} u^\mu
- \frac{c_1}{2\Theta} q^\alpha \Pi_\alpha^\mu. \label{eq:entropyExpansionIS}
\end{equation}
This is precisely what is done in Israel-Stewart theory, except that our bulk viscosity is $0$ (and working in the Eckart frame eliminates another term related to mass diffusion).  The factors of $1/\Theta$ are chosen for convenience. { The ordering in the above expansion is $|q^\mu| \sim |\Pi^{\mu \nu}| \sim \epsilon \ll 1 $. The first term $s\rho u^\mu$ is the leading order term $O(\epsilon^0)$ and is present even in the ideal case. The term $\propto q^\mu$ is first order in a dissipative field $O(\epsilon)$, i.e. here the heat flux $q^\mu$. The terms $\propto q^\alpha q_\alpha$, $\Pi^{\alpha \beta} \Pi_{\alpha \beta}$ and $q^\alpha \Pi^{\mu}_\alpha$ are second order $O(\epsilon^2)$. Note that there is no viscosity contribution at first order in the above expansion, as is explained in section (\ref{subsectionEntropyCurrent}).}

We now set $c_1 = 0$ to simplify the model further.  The neglected term couples $q$ and $\Delta P$. The value of $c_1$ cannot be determined at this thermodynamic level and one has to resort to kinetic theory (see section C in \citealt{Bouras2010}). However, its choice does not affect the amount of entropy production. We remark further on the effect of $c_1 \neq 0$ at the end of the derivation. { Thus we have
\begin{equation}
s^\mu = s \rho u^\mu 
+ \frac{a_1}{\Theta} q\,\hat{b}^\mu 
- \frac{b_1}{2\Theta} q^2 u^\mu
- \frac{b_2}{3\Theta} \Delta P^2 u^\mu . \label{eq:entropyExpansion}
\end{equation}
}
{ where we have used  $\Pi^{\mu\nu} \Pi_{\mu\nu} = \frac{2}{3}\Delta P^2$}. Now evaluate $\nabla_\mu s^\mu$.  First,
\begin{equation}
\nabla_\mu (s \rho u^\mu) = \frac{1}{\Theta} u_\nu \nabla_\mu T^{\mu\nu}_{C+V} \label{eq:divOfEntropyZerothOrderTerm}
\end{equation}
where $T_{C+V}$ is the sum of the conduction and viscosity terms in the stress-energy tensor.  Then the conduction terms give
\begin{equation}
u_\nu \nabla_\mu (u^\mu q^\nu + q^\mu u^\nu) = - q^\mu a_\mu - \nabla_\mu q^\mu.
\label{eq:derivOfTCond}
\end{equation}
where $a_\mu \equiv u^\alpha \nabla_\alpha u_\mu$ is the four-acceleration, { and we have used the constraint $u_\mu q^\mu = 0$}. The viscosity terms give
\begin{equation}
u_\mu \nabla_\nu ( -\Delta P (\bh^\mu\bh^\nu -\frac{1}{3} h^{\mu\nu})) = \Delta P \left(\bh^\mu \bh^\nu \nabla_\mu u_\nu - \frac{1}{3} \nabla_\mu u^\mu\right).
\end{equation}
{ In deriving the above, we have used the constraint $u_\mu \hat{b}^\mu = 0 \Rightarrow u_\mu \hat{b}^\nu \nabla_\nu \hat{b}^\mu = -\hat{b}^\mu \hat{b}^\nu \nabla_\nu u_\mu$.}
Next, the first order term in $q$ is
\begin{equation}
\nabla_\mu \left(\frac{a_1}{\Theta} q^\mu \right) =
\frac{a_1}{\Theta} \nabla_\mu q^\mu + q^\mu \nabla_\mu \left(\frac{a_1}{\Theta}\right) 
\end{equation}
the second order term in $q$ is
\begin{equation}
-\nabla_\mu \left( \frac{b_1}{2 \Theta} q^2 u^\mu \right) =
- \frac{b_1 q}{\Theta} \frac{dq}{d\tau} - \frac{q^2}{2} \nabla_\mu \left( \frac{b_1 u^\mu}{\Theta}\right)
\end{equation}
where $dq/d\tau = u^\mu \nabla_\mu q$, and the second order term in $\Delta P$ is 
\begin{equation}
-\nabla_\mu \left( \frac{b_2}{3 \Theta} \Delta P^2 u^\mu\right) = 
- \frac{2 b_2 \Delta P}{3 \Theta} \frac{d\Delta P}{d\tau} - \frac{\Delta P^2}{3}\nabla_\mu \left( \frac{b_2 u^\mu}{\Theta}\right)
\end{equation}

Assembling all the first-order terms in $q$ from $\nabla_\mu s^\mu$,
\begin{equation}
\frac{a_1 - 1}{\Theta} \nabla_\mu q^\mu - q^\mu a_\mu + q^\mu \nabla_\mu \left(\frac{a_1}{\Theta}\right).
\end{equation}
The term proportional to $\nabla_\mu q^\mu$ has indeterminate sign, so we choose $a_1 = 1$. 
Gathering all the terms { (first order + second order)} in $q$ (and writing down only the $q$ terms of $\nabla_\mu \left(s\rho u^\mu\right)$), we find
\begin{widetext}
\begin{equation}
{ \nabla_\mu \left(s\rho u^\mu + \frac{q}{\Theta} \hat{b}^\mu - \frac{b_1}{2\Theta}q^2 u^\mu\right)} = q^\mu \left[ -\frac{\nabla_\mu \Theta}{\Theta^2} - \frac{a_\mu}{\Theta} - \frac{q_\mu}{2}\nabla_\alpha \left(\frac{b_1 u^\alpha}{\Theta}\right) - \frac{b_1 \bh_\mu}{\Theta}\frac{dq}{d\tau}\right].
\end{equation}
\end{widetext}
Evidently this can be positive definite if the quantity in square brackets $ = \beta_1 q_\mu$ and $\beta_1 > 0$.  Remarkably, by applying this condition we find an evolution equation for $q$:
\begin{equation}
\frac{d q}{d\tau} = -\frac{\Theta}{b_1} \left(
\beta_1 q + \frac{\bh^\mu (\nabla_\mu \Theta + \Theta a_\mu)}{\Theta^2} + q\nabla_\alpha \left( \frac{b_1 u^\alpha}{2 \Theta}\right)\right)
\end{equation}
All that remains is to fix the constants $b_1$ and $\beta_1$.  For $\beta_1$, we require that $q$ asymptote to its non-relativistic value, $-\rho\chi \grad \Theta$, where $\chi$ is the conductive diffusivity and has dimensions of a length times a velocity.  Then $\beta_1 = (\rho\chi \Theta^2)^{-1}$.  For $b_1$, we require that it be proportional to a relaxation timescale $\tau_R$.  Then  $b_1 = \tau_R/(\rho\chi \Theta)$.  Gathering all together, our final evolution equation for the heat flux is
\begin{equation}
\frac{dq}{d\tau} = -\frac{q - q_0}{\tau_R} - \frac{q}{2} \,  \frac{d}{d\tau} \log\left(\frac{\tau_R}{\chi P^2}\right)
\label{eq:qevol}
\end{equation}
where 
\begin{equation}
q_0 \equiv -\rho\chi\bh^\mu (\nabla_\mu \Theta + \Theta a_\mu) \label{eq:q0def}
\end{equation}
is the classical Eckart heat flux projected onto the magnetic field.   { The heat flux $q$ relaxes to the first order (Eckart) heat flux $q_0$ over a timescale $\tau_R$. The additional term on the right is a second order correction and is formally necessary to ensure the positivity of entropy production. The importance of this term in a full calculation can only be gauged by performing a calculation with and without this term.} Notice that equation (\ref{eq:qevol}) can be rewritten in the remarkably simple, scaled form
\begin{equation}
\frac{d Q}{d\tau} = - \frac{Q - Q_0}{\tau_R},
\label{eq:simpqevol}
\end{equation}
with $Q \equiv q (\tau_R/(\chi P^2))^{1/2}$.

Next, assemble all terms in $\nabla_\mu s^\mu$ depending on $\Delta P$ (and writing down only the $\Delta P$ terms of $\nabla_\mu \left(s\rho u^\mu\right)$) to find
\begin{widetext}
\begin{equation}
{
\nabla_\mu \left(s\rho u^\mu - \frac{b_2}{3\Theta} \Delta P^2 u^\mu\right)} =
\Delta P \left[
\frac{1}{\Theta}\left(\bh^\mu \bh^\nu \nabla_\mu u_\nu - \frac{1}{3} \nabla_\mu u^\mu\right) - \frac{\Delta P}{3} \nabla_\mu \left(\frac{b_2 u^\mu}{\Theta} \right) - \frac{2 b_2}{3 \Theta} \frac{d\Delta P}{d\tau}\right].
\end{equation}
\end{widetext}
Evidently this will be positive definite if the quantity in square brackets $ = \beta_2 \Delta P$ and $\beta_2 > 0$.  This provides an evolution equation for $\Delta P$:
\begin{widetext}
\begin{equation}
\frac{d\Delta P}{d\tau} = \frac{3 \Theta}{2 b_2} \left(-\beta_2 \Delta P + \frac{1}{\Theta}(\bh^\mu \bh^\nu \nabla_\mu u_\nu - \frac{1}{3} \nabla_\mu u^\mu)\right) - \frac{\Theta \Delta P}{2 b_2} \nabla_\mu \left(\frac{b_2 u^\mu}{\Theta}\right).
\end{equation}
\end{widetext}
Again set the coefficients by requiring that $b_2$ be proportional to a relaxation time $\tau_R$ and that $\Delta P$ asymptote to its classical non-relativistic limit $= 3 \rho \nu (\bhat \bhat : \grad \bv - \frac{1}{3} \grad \cdot \bv)$ (see the Appendix), with $\nu \equiv$ kinematic viscosity, to find $b_2 = \tau_R/(2 \rho \nu)$ and $\beta_2 = (3\rho\nu \Theta)^{-1}$.  Gathering all together, the final evolution equation for $\Delta P$ is
\begin{equation}
\frac{d\Delta P}{d\tau} = - \frac{\Delta P - \Delta P_0}{\tau_R} - \frac{\Delta P}{2}\frac{d}{d\tau}\log \left(\frac{\tau_R}{\rho\nu P}\right),
\label{eq:dPevol}
\end{equation}
where
\begin{equation}
\Delta P_0 \equiv 3\rho \nu (\bh^\mu \bh^\nu \nabla_\mu u_\mu - \frac{1}{3} \nabla_\mu u^\mu)
\label{eq:dP0def}
\end{equation}
is a covariant generalization of the \cite{Braginskii1965} model, in which collisions balance the forcing of anisotropy by the velocity field (see appendix A). { As we already saw in (\ref{eq:qevol}) where the heat flux relaxes to Eckart theory over a timescale $\tau_R$, the pressure anisotropy $\Delta P$ also, relaxes to its first order value $\Delta P_0$, with the second term on the right hand side of (\ref{eq:dPevol}) being a higher order correction required for positivity of the entropy production.} Notice that equation (\ref{eq:dPevol}) can also be written in simplified, scaled form:
\begin{equation}
\frac{d D}{d\tau} = -\frac{D - D_0}{\tau_R}
\end{equation}
where $D \equiv \Delta P (\tau_R/(\rho\nu P))^{1/2}$.

It is also useful to gather the full, relativistic dissipation function:
\begin{widetext}
\begin{equation} \label{eq:entropyProduction}
\nabla_\mu s^\mu { = \nabla_\mu \left(\rho s u^\mu + \frac{q}{\Theta} \hat{b}^\mu - \frac{\tau_R}{2\rho \chi \Theta^2} q^2 u^\mu - \frac{\tau_R}{6\rho \nu \Theta} \Delta P^2 u^\mu \right)} = \frac{q^2}{\chi \rho \Theta^2} + \frac{1}{3}\frac{\Delta P^2}{\nu \rho \Theta}
\end{equation}
\end{widetext}
which is positive definite, has the correct units, and reduces to the correct dissipation function in the non-relativistic limit { as we shall show in the next section. Note that the right hand side of (\ref{eq:entropyProduction}) is a function of the full heat flux $q$ and the pressure anisotropy $\Delta P$, that are solved for using (\ref{eq:qevol}) and (\ref{eq:dPevol}) respectively, and not the relaxed forms $q_0$ and $\Delta P_0$.} The above value of entropy production is invariant under the choice of the cross coupling coefficient $c_1$ in (\ref{eq:entropyExpansion}). However, had we not set $c_1=0$, the evolution equations for $q$ (\ref{eq:qevol}) and $\Delta P$ (\ref{eq:dPevol}) would both have additional terms which couple $q$ and $\Delta P$ to each other. \footnote{{ The additional terms when the cross coupling coefficient $c_1 \neq 0$ are of the form $\frac{dq}{d\tau} = ... + (...)\hat{b}^\mu \nabla_\mu \Delta P + (...)\Delta P\nabla_\mu \hat{b}^\mu$ and $\frac{d\Delta P}{d\tau} = ... + (...)\hat{b}^\mu \nabla_\mu q + (...)q\nabla_\mu \hat{b}^\mu$. Therefore, $q$ and $\Delta P$ are driven not only by gradients of the background thermodynamic quantities $\hat{b}^\mu \nabla_\mu \Theta$ and $\hat{b}^\mu \hat{b}^\nu \nabla_\mu u_\nu$, but by gradients of each other. Such terms become important in the collisionless limit. The cross coupling coefficients could potentially be used to derive a more accurate model where the dissipative fields are $q_\parallel$, $q_\perp$, $\Delta P_\parallel$ and $\Delta P_\perp$, with the correct coupling between the fields. We leave this to future work. \label{footnote:crosscoupling} }}

The model given by equations (\ref{eq:qevol}) and (\ref{eq:dPevol}) is derived using precisely the same procedure as the Israel-Stewart model, but the complexity is greatly reduced because -- thanks to the magnetic field -- there are only two nonideal degrees of freedom.  We will also see below that the model is not subject to the linear instabilities of the isotropic first order theory (where $b_1 = b_2 = 0$) discovered by \cite{Hiscock1985}, provided the damping timescale $\tau_R$ is chosen appropriately, as found by \cite{Hiscock1983} and as we shall see in Sec.~\ref{sec:linear}.  

{
\section{Connection to non-relativistic dissipative theory}}
In this section we compare the equations governing the entropy scalar $s$ and the entropy current $s^\mu$ (\ref{eq:entropyProduction}) to their non-relativistic counterparts. It is important to stress that the entropy scalar $s$ and the entropy current $s^\mu$ are distinct quantities which obey separate evolution equations. Both of these have an analog in non-relativistic dissipative hydrodynamics. In equilibrium, the two quantities are related by $s = -s^\mu u_\mu/\rho$, but this is not true in general. This can be seen explicitly in (\ref{eq:entropyExpansion}) where there are second order differences ($\sim q^2, \Delta P^2)$ between $s$ and $-s^\mu u_\mu / \rho$.

{ \subsection{Entropy scalar $s$} }
In non-relativistic dissipative hydrodynamics, the evolution equation for the entropy scalar $s$ (\citealt{Landau1987}) is
\beqn
\rho \Theta \frac{Ds}{Dt} = -\grad \cdot \mathbf{q} - \mathbf{\Pi} : \grad \mathbf{u} \label{eq:nonreldsdtau}
\eeqn
where $D/Dt$ is the convective derivative $\partial/\partial t + \mathbf{u}\cdot\grad$. We now derive the corresponding equation in relativistic dissipative hydrodynamics starting from (\ref{eq:dsdtau}), which itself has been derived from the first law of thermodynamics. Proceeding to do so
\beqn
\rho \Theta \frac{ds}{d\tau} & = & -u_\nu \nabla_\mu T^{\mu \nu}_{\mathrm{ideal}}  =  u_\nu \nabla_\mu T^{\mu \nu}_{C+V} \\
& = & u_\nu \nabla_\mu\left(q^\mu u^\nu + q^\nu u^\mu + \Pi^{\mu\nu}\right)\\
& = & -\nabla_\mu q^\mu -q^\mu a_\mu - \Pi^{\mu \nu} \nabla_\mu u_\nu \label{eq:reldstau}
\eeqn
where we have used (\ref{eq:derivOfTCond}) and the constraint $\Pi^{\mu \nu}u_\nu=0 \Rightarrow u_\nu\nabla_\mu\Pi^{\mu\nu} = -\Pi^{\mu\nu}\nabla_\mu u_\nu$. The above equation is independent of the model for $q^\mu$ and $\Pi^{\mu \nu}$, and only uses the form of the dissipative component $T_{C+V}^{\mu\nu}$ of the stress-tensor. Thus, the equation is valid for both the first order Eckart theory as well as the second order Israel-Stewart theory. The difference between the non-relativistic equation (\ref{eq:nonreldsdtau}) and the relativistic equation (\ref{eq:reldstau}), apart from the 3-derivatives transforming into covariant derivatives, is the presence of the $q^\mu a_\mu$ term, which has no equivalent in the non-relativistic case.

{ \subsection{Entropy current $s^\mu$}  \label{subsectionEntropyCurrent} }
The equation for the divergence of the entropy current (\ref{eq:entropyProduction}) has been derived using a second order ansatz for the entropy current (\ref{eq:entropyExpansion}). However, the value of the entropy production is the same as in first order theories, two of which are the relativistic theory by Eckart, and the classical non-relativistic theory of dissipation. Below we show, starting from the evolution equation for the entropy scalar $s$ from the previous subsection, that the entropy production in (\ref{eq:entropyProduction}) is true for both the relativistic and non-relativistic first order theories.

\begin{itemize}
\item
For the relativistic first order Eckart theory, we start with (\ref{eq:reldstau}) and use the first order isotropic heat flux $q^\mu = q_0^\mu = -\rho \chi h^{\mu \nu}\left(\nabla_\mu \Theta + \Theta a_\mu \right)$, the first order shear tensor $\Pi_{\mu \nu} = \Pi_{0\mu \nu} = -2 \rho \nu \langle \nabla^\mu u^\nu \rangle = -\rho \nu h^\alpha_\mu h^\beta_\nu\left(\nabla_\alpha u_\beta + \nabla_\beta u_\alpha - (2/3)h_{\alpha \beta}h^{\kappa \eta}\nabla_\kappa u_\eta \right)$ (\citealt{Andersson2007}), along with the relativistic continuity equation (\ref{eq:continuity}) to get
\beqn
\nabla_\mu \left(\rho s u^\mu + \frac{q_0^\mu}{\Theta}\right) & = &  \frac{q_0^\mu q_{0\mu}}{\rho \chi \Theta^2} + \frac{\Pi_0^{\mu \nu} \Pi_{0\mu \nu}}{2 \rho \nu \Theta}
\label{eq:entropyProductionEckart}
\eeqn
In the anisotropic case, we again start with (\ref{eq:reldstau}), but now use the heat flux $q^\mu = q_0\; \hat{b}^\mu$ and the shear stress $\Pi^{\mu \nu} = -\Delta P_0 \left(\hat{b}^\mu \hat{b}^\nu - \frac{1}{3}h^{\mu\nu} \right)$, where $q_0$ and $\Delta P_0$ are given by (\ref{eq:q0def}) and (\ref{eq:dP0def}) respectively, along with (\ref{eq:continuity}) to get
\beqn
\nabla_\mu \left(\rho s u^\mu + \frac{q_0}{\Theta} \hat{b}^\mu\right) & = &  \frac{q_0^2}{\rho \chi \Theta^2} + \frac{\Delta P_0^2}{3 \rho \nu \Theta} \label{eq:entropyProductionEckartAnisotropic}
\eeqn

We see that the first order Eckart theory satisfies (\ref{eq:entropyProduction}) with $s^\mu = \rho s u^\mu + q_0^\mu/\Theta$, and has the same amount of entropy production as the second order Israel-Stewart theory. However, in this case the entropy current is sourced by $q_0$ (\ref{eq:q0def}) and $\Delta P_0$ (\ref{eq:dP0def}) which are directly related to gradients of thermodynamic quantities, as opposed to $q$ and $\Delta P$ on the right hand side of (\ref{eq:entropyProduction}), which are solved for using (\ref{eq:qevol}) and (\ref{eq:dPevol}) respectively. It can also be seen from the above why the viscous stress $\Pi^{\mu \nu}$ does not contribute at first order in the entropy expansion, since $\Pi_{\mu \nu} = \Pi_{0\mu \nu} = -2\rho \nu \langle\nabla_\mu u_\nu \rangle$ produces the right amount of dissipation $\Pi_{0\mu \nu} \Pi_0^{\mu \nu}/(2 \rho \nu \Theta)$ on the right hand side, without any corresponding term in the entropy current on the left hand side of (\ref{eq:entropyProductionEckart}).

\item
In the non-relativistic case, we start from (\ref{eq:nonreldsdtau}) and use the classical heat flux $\mathbf{q} = -\rho \chi \grad \Theta$, the shear stress $\mathbf{\Pi} = -2 \rho \nu \langle \grad \mathbf{u} \rangle$, which in component form is explicitly $\Pi_{ik} = -\rho \nu \left(\partial v_i/\partial x_k + \partial v_k/\partial x_i - (2/3)\delta_{ik}\partial v_l/\partial x_l \right)$ (\citealt{Landau1987}), and the non-relativistic continuity equation to get
\beqn
\frac{\partial (\rho s)}{\partial t} + \grad\cdot\left(\rho s \mathbf{u} + \frac{\mathbf{q}}{\Theta}\right) & = & \frac{\mathbf{q}\cdot\mathbf{q}}{\rho \chi \Theta^2} + \frac{\mathbf{\Pi} : \mathbf{\Pi}}{2 \rho \nu \Theta} \label{eq:nonrelEntropyProduction}
\eeqn
The above is true for the anisotropic case as well, where we have the heat flux $\mathbf{q} = -\rho \chi \hat{\mathbf{b}} \hat{\mathbf{b}}\cdot \grad \Theta$, the shear stress $\mathbf{\Pi} = -\Delta P_0
\left[\bhat\bhat - \mathbf{I}/3\right]$, and $\Delta P_0 = 3\rho \nu\left[\bhat\bhat:\grad \boldsymbol{v}
-\grad\cdot\boldsymbol{v}/3\right]$ (\citealt{Braginskii1965}). Clearly, (\ref{eq:nonrelEntropyProduction}) is the non-relativistic limit of (\ref{eq:entropyProductionEckart}), where $s^0 \rightarrow \rho s$ when $v/c \ll 1$, since in this limit, $u^0 \rightarrow 1$ and $q^0 \rightarrow 0$.
\end{itemize}
Thus, the value of entropy production (right hand side of (\ref{eq:entropyProduction})), which is second order in $q^\mu$ and $\Pi^{\mu \nu}$, is the same for the entropy current expanded to first order (relativistic and non-relativistic), as well as to second order. However, expanding the entropy current to third order leads to additional higher order terms on the right hand side of (\ref{eq:entropyProduction}) (\citealt{El2010}).

\section{Model Parameters: Viscosity, Conductivity, and Relaxation Time}
\label{sec:modparam}

The parameters of our theory are $\tau_R, \nu,$ and $\chi$. { The entropy production (\ref{eq:entropyProduction}) can be interpreted in two ways, depending on whether the plasma is collisional/weakly collisional or collisionless. If the plasma is collisional/weakly collisional, it is the microscopic entropy production due to Coulomb scatterings and $\tau_R$ is the Coulomb scattering time scale, i.e., the mean free time between particle-particle collisions. The transport coefficients $\nu$, and $\chi$ are set by this time scale $\tau_R$. In particular, both $\nu$ and $\chi$ are of order $c_s^2 \tau_R$ in relativistic collisional kinetic theory:}
\begin{equation}
\chi = \phi c_s^2 \tau_R \hspace{2cm} {\rm and} \hspace{2cm} \nu = \psi c_s^2 h \tau_R
\label{eq:chi-nu}
\end{equation}
where $\phi$ and $\psi$ are constant dimensionless parameters and $h \equiv 1 + \gamma u/(\rho c^2)$ is the relativistic enthalpy.

For a monoatomic ideal gas in the Chapman-Enskog theory $\phi = (15/4)\psi$.  In the non-relativistic Braginskii theory, where Coulomb interactions dominate, $\phi \simeq 4.1\psi$ \citep[see][]{Kulsrud2005}.  For a relativistic hard sphere gas \citep[see the clear discussion of][]{Cercignani2001} $\nu \propto T$ for $\Theta \equiv k T/(m c^2) \gg 1$ ($m \equiv$ molecular weight).

{ In a collisionless plasma, because of the absence of Coulomb scatterings, the scattering time scale diverges $\tau_R \rightarrow \infty$ and the entropy does not increase. In this case, (\ref{eq:entropyProduction}) is the increase in a \emph{coarse grained} entropy and $\tau_R$ is the mean free time between wave-particle scatterings. This is because coarse graining the Vlasov equation leads to a Fokker-Planck equation with an effective collision operator, as is done in quasi-linear theory. Therefore, we still use the closures (\ref{eq:chi-nu}) but with a different interpretation of $\tau_R$ compared to the collisional case.} Wave-particle scattering is determined primarily by fluctuations in the electromagnetic field that have frequencies of order the cyclotron frequency of the particles of interest (ions, electrons) \citep{Kulsrud2005}. Such fluctuations can either be produced by a  cascade from larger scales or by velocity space instabilities that directly excite high frequency fluctuations.  Unfortunately, the efficiency of wave-particle scattering by these processes is  not fully understood.  Moreover, the high frequency turbulent fluctuations that dominate scattering cannot be resolved in fluid (GRMHD) simulations, so subgrid models of the scattering rate are necessary.   The formalism developed in the previous sections is sufficiently general that as the theoretical understanding of wave particle scattering develops, increasingly sophisticated models of $\tau_R$ can be implemented in our model.   In particular, we stress that the relaxation time, viscosity, and conductivity in our model can be functions of the local plasma conditions (including, e.g., the local plasma $\beta$, the amplitude of the turbulent fluctuations, etc.). Here we provide a rough guide to some of the key physics that motivates particular choices for $\tau_R$, $\nu$, and $\chi$.

It is unclear to what extent magnetized turbulence in accretion disks (driven by the magnetorotational instability) produces significant wave-particle scattering.  The energetically dominant component of the turbulence  (associated with the slow and Alfv\'en modes)  does not produce efficient scattering because it does not cascade to high frequencies \citep{GS1995, Quataert1998}.  The fast mode component can in principle cascade to high frequencies \citep{Yan2004}, but is very strongly damped in low-collisionality plasmas, which likely limits its importance.   Absent significant scattering by the high frequency tail of a turbulent cascade, the most important source of scattering is due to velocity space instabilities. These are instabilities in which temperature variations in different directions or local streaming velocities relative to the magnetic field frame provide a source of free energy to drive instabilities (e.g., \citealt{Stix1992}).  Key examples include the firehose, mirror, and ion cyclotron instabilities for the ions and the electron firehose and whistler instabilities for the electrons (we show in \S \ref{sec:linear} that the firehose instability is captured by the fluid model developed in this paper; this is not true for resonant instabilities such as the ion cyclotron and whistler instabilities).  { The saturation of these instabilities is an active area of research with implications for galaxy cluster plasmas and the solar wind, as well as accretion disks (e.g., \citealt{Kunz2014,Riquelme2015,Sironi2015,Hellinger2015})}.   Note that for the ions in RIAFs, a non-relativistic theory ($\theta_i \equiv k T_i/(m_i c^2) \lesssim 1$) is sufficient, since $\theta_i \sim (H/r)^2 (G M/(r c^2))^{1/2}$, and $r > H \equiv$ scale height.  However, for the electrons a relativistic theory is required since in the RIAF model $\theta_e \gtrsim 1$ and indeed relativistic electrons are observed close to Sgr A* \citep{Doeleman2008}.

Velocity space instabilities in low collisionality plasmas such as RIAFs are believed to arise because shearing, heating, expansion, or compression drives the plasma towards an unstable configuration, initiating the instability.  The scattering rate produced once the velocity space instabilities saturate is of the order of the timescale on which the instabilities are driven (see \citealt{Kunz2014,Riquelme2015}).  This balance between driving and scattering timescales maintains the plasma near the marginal state for the instability.  These considerations provide good motivation for one specific choice of the relaxation time in low-collisionalty plasmas: $\tau_R$ of the order of the dynamical time $\tau_d$, because the latter sets the characteristic timescale for heating, expansion, etc.   For black hole accretion problems, this suggests $\tau_R \sim \tau_d \simeq (G M/r^3)^{-1/2}$.   

More specifically, however, velocity space instabilities will only set in if the free energy driving the instability is sufficiently large.  For example, the heat flux due to conduction cannot exceed the saturated value of $q \simeq \Phi \rho c_s^3$, where $\Phi$ is a dimensionless constant of order unity, but the electron whistler instability implies an even more stringent limit on the heat flux for high $\beta$ plasmas \citep{Pistinner1998}.   As a second example, a non-relativistic plasma is firehose (mirror) unstable if $\Delta P/P \le -2/\beta$ ($\Delta P/P \gtrsim 1/\beta$).  Observations of the solar wind show that the plasma pressure anisotropy obeys these constraints to reasonable accuracy (e.g., \citealt{Hellinger2006,Bale2009}).  Thus a physical model of accretion disk viscosity should ensure that $|\Delta P|$ does not exceed these bounds. { Since, both viscosity and thermal conductivity are related to  the relaxation time scale $\tau_R$ in (\ref{eq:chi-nu}), this implies a modification of the thermal conductivity as well, when $\tau_R$ is decreased.}

One strategy for reducing $\tau_R$ in the presence of a small scale instability is to set
\begin{equation}
\tau_R = \tau_d f\left(\frac{x}{x_{crit}} \right) \label{eq:taumod}
\end{equation}
where $x$ is some parameter (e.g. $\Delta P$) that has a critical value $x_{crit}$ for instability.  The function $f$ is arbitrary but should have (1) $f(0) = 1$; (2) $f'(0) = 0$; (3) $f(1) \ll 1$.  One function with the desired properties for $f$ is the Fermi-Dirac distribution:
\begin{equation}
f(y) = \frac{1}{e^{(y-1)/\lambda} + 1} \label{eq:regulator}
\end{equation}
for $y > 0$, and $f(y) = 1$ for $y < 0$.  $\lambda$ is an adjustable parameter that determines the width of the transition to small $\tau_R$.  As an example, saturated conduction can be implemented by sending $\tau_R \rightarrow \tau_d f(q/q_{crit})$, where $q_{crit} = \Phi \rho c_s^3$ is the maximum heat flux (and we have assumed a characteristic, unsaturated, relaxation time of $\tau_d$)\footnote{{Saturated conduction is technically a collisionless effect and in this limit, the anisotropic pressure depends not only on the shear projected onto the field lines, but also on the gradients of the heat flux which we have ignored. See also footnote \ref{footnote:crosscoupling}.}}. { To consider several instabilities with instability threshold ratios $y_1$, $y_2$, $y_3...$, one can set $\tau_R = \tau_df(y_1)f(y_2)f(y_3)...$}

{
To understand the effect of (\ref{eq:taumod}) and the associated reduction in the relaxation time near an instability threshold, consider an instability with threshold pressure anisotropy $\Delta P_{crit}$. As $\Delta P \rightarrow \Delta P_{crit}$, the relaxation time is reduced.  Because $\Delta P_0 \propto \nu \propto \tau_R$, this reduces both $\Delta P_0$ and the time required for $\Delta P$ to relax to $\Delta P_0$. Thus the plasma quickly falls below the instability threshold. Once below the threshold, $\tau_R$ becomes $\tau_d$ again, and $\Delta P$ relaxes to the large $\Delta P_0$ over a dynamical time scale. When the threshold is crossed again, (\ref{eq:taumod}) takes effect and the cycle restarts. The net effect of this cycle is to set up a feedback loop that results in $q$ and $\Delta P$ being pinned to their values at an instability threshold, in a statistical sense. At every instant, the process generates entropy according to (\ref{eq:entropyProduction}), which only involves $q$ and $\Delta P$ and not $q_0$ and $\Delta P_0$.
}

\section{Linear Theory}
\label{sec:linear}

Here we address the following questions: (1) what are the characteristic speeds in our extended MHD model? These are needed to determine the Courant condition in explicit numerical evolution; (2) which choices of model parameters $\tau_R, \chi, \nu$ yield a stable model when the initial state is an equilibrium?  (3) under what conditions does one recover the firehose instability of an initially anisotropic state? (4) how large a heat flux can the model admit before it loses its stability and hyperbolicity? (5) what is the stability of the model in a frame not comoving with the fluid? and finally (6) under what conditions is the model causal?

Consider an initially homogeneous magnetized fluid in Minkowski space.  The initial state in the fluid frame has $u^\mu = \{1,0,0,0\},$ $\rho = \rho_0$, $u = u_0$, $b^\mu = (0, b \sin\theta, 0, b \cos\theta)$, and $q = \Delta P = 0$.  The initial temperature is is $T_0 = P_0/\rho_0 = (\gamma - 1) u_0/\rho_0$.  We perturb around this initial state, e.g. $q \rightarrow 0 + \delta q$ with $\delta q \propto \exp(i k x - i\omega t)$, linearize, and find the dispersion relation $D(\omega,k) = 0$. 

It is worth first revisiting the linear theory for relativistic ideal MHD.  As usual, the Alfv\'en waves factor and the Alfv\'en velocity is
\begin{equation}
v_A^2 = \frac{b^2}{\rho_0 h_0 + b^2}.
\end{equation}
The slow and fast modes combine in a complicated, fourth-order dispersion relation.  The special cases parallel and perpendicular to the field give the sound speed
\begin{equation}
c_s^2 = \frac{\gamma P_0}{\rho_0 h_0}
\end{equation}
and the fast magnetosonic speed
\begin{equation}
v_M^2 = \frac{b^2 + \gamma P_0}{\rho_0 h_0 + b^2}
\end{equation}
respectively.  

A rigorous stability analysis would require a study of the general, ninth-order dispersion relation.  Instead we consider the special cases of propagation parallel and perpendicular to the magnetic field lines for conduction only and for viscosity only, and analyze the resulting dispersion relations.

{\em $\nu = 0$, $k \parallel B$.}  The four Alfv\'en modes are unaffected, as one might expect because they do not perturb the temperature.  The remaining four modes arise from the entropy mode and the sound waves coupled to evolution of $\delta q$.  The dispersion relation is:
\begin{widetext}
\begin{equation}
(\omega^2 - c_s^2 k^2) \omega (\omega \tau_R + i) -
\frac{\chi c_s^2}{\gamma c^4}\left( (\omega^2 - (\gamma - 1) c^2 k^2)^2 + c^2 k^2 (\omega^2 - c_s^2 k^2) \frac{\gamma(\gamma - 1) c^2}{c_s^2} \right)
 = 0.
\label{eq:dispql}
\end{equation}
\end{widetext}
Evidently, the sound waves emerge in the ideal limit.

{\em $\nu = 0$, $k \perp B$.}  We recover five zero frequency modes, two magnetosonic modes, and one new mode:
\begin{equation}
\omega = -i \left(\tau_R - \frac{\chi c_s^2}{\gamma c^4}\right)^{-1}.
\label{eq:dispqp}
\end{equation}

{\em $\chi = 0$, $k \parallel B$.} We recover four Alfv\'enic modes (propagating in each direction with two polarizations), a zero frequency mode that asymptotes to the entropy mode, and modes that couple sound waves and $\Delta P$:
\begin{equation}
\omega^3 
+ i \omega^2 \frac{1}{\tau_R}
-  \omega \left( \frac{4 \nu k^2}{3\tau_R (1 + \gamma u_0/\rho_0 c^2)} 
+ c_s^2 k^2 \right)
- i \frac{c_s^2 k^2}{\tau_R} = 0.
\label{eq:dispdPl}
\end{equation}

{\em $\chi = 0$, $k \perp B$.} We recover a damped viscous mode coupled to the two magnetosonic modes:
\begin{equation}
\omega^3
+ i\omega^2 \frac{1}{\tau_R}
- \omega \left(\frac{\nu k^2 \rho_0 c^2}{3 \tau_R (b^2 + \rho_0 c^2 + \gamma u_0)} + k^2 v_M^2 \right)
- i \frac{k^2 v_M^2}{\tau_R} = 0.
\label{eq:dispdPp}
\end{equation}

\subsection{Characteristic Speeds}

The introduction of evolution equations for $q$ and $\Delta P$ makes the equations hyperbolic and therefore also introduces new characteristic speeds.  On dimensional grounds we expect a characteristic speed associated with conduction $v_q^2 \propto \chi/\tau_R$ and a characteristic speed associated with viscosity $v_{\Delta P}^2 \propto \nu/\tau_R$. What are the constants of proportionality?

First consider the pure viscosity case in the limit $k \rightarrow \infty$.  For parallel propagation
\begin{equation}
\omega^2 = k^2 \left( c_s^2 + \frac{4}{3} \frac{\nu}{h_0 \tau_R}
\label{eq:dispnul}
\right)
\end{equation}
and for perpendicular propagation
\begin{equation}
\omega^2 = k^2 \left( v_M^2 + \frac{1}{3} \frac{\nu}{\tau_R (1 + \gamma u_0/\rho_0 + b^2/\rho_0)}
\right).
\end{equation}
This motivates the definition 
\begin{equation}
v_{\Delta P}^2 \equiv \frac{4}{3}\frac{\nu}{h_0 \tau_R}
\end{equation}
as a characteristic viscous speed.  In our closure model $v_{\Delta P} = c_s \sqrt{4\psi/3}$, and in a numerical implementation the Courant condition needs to be adjusted accordingly.

Next consider the pure conduction case in the limit $k \rightarrow \infty$.  For parallel propagation the general solution for $\omega^2$ for the coupled entropy-conduction-sound wave is complicated and not much simpler than the dispersion relation itself, but the expression
\begin{equation}
\omega^2 = \frac{1}{2} k^2 \left( c_s^2 +  v_q^2 \pm (c_s^4 + v_q^4)^{1/2} \right)
\end{equation}
provides a reasonable first approximation (although it does not reveal that $\omega^2 \propto 1/(\gamma(\gamma - 1) c^4 - c_s^2 v_q^2)$, which leads to slightly higher signal speeds when $c_s, v_q \sim c$).  Here 
\begin{equation}
v_q^2 \equiv (\gamma - 1) \frac{\chi}{\tau_R}
\end{equation}
is a characteristic conduction speed.  For perpendicular propagation, the excitation of $q$ is damped and nonpropagating.
In our closure model $v_q = c_s \sqrt{\phi(\gamma-1)}$ and so in a numerical implementation of the model the Courant condition must be adjusted accordingly.  

\subsection{Stability}

It is well known that the Eckart first order conduction model \citep[see][p. 567]{Misner1973} is subject to a dramatic instability with growth rate $\propto 1/\chi$ \citep{Hiscock1985}.  The Eckart theory is recovered in the limit $\tau_R \rightarrow 0$ in our model, so we expect that it will be unstable if we set $\tau_R$ too small.

To make this idea more precise, consider the $k \rightarrow 0$ limit of (\ref{eq:dispql}).  We find
\begin{equation}
\omega = -i \left(\tau_R - \frac{\chi c_s^2}{{\gamma c^4}}\right)^{-1}
\end{equation}
where we have temporarily restored factors of $c$ and $c_s^2 = \gamma P_0/(\rho_0 + \gamma u_0/c^2)$. Evidently the model is unstable if 
\begin{equation}
\tau_R < \chi c_s^2/(\gamma c^4)
\end{equation}
which is the same instability condition found by \cite{Hiscock1985} for an isotropic viscosity.  For propagation perpendicular to the field, (\ref{eq:dispqp}), we recover the same stability condition. Setting $\chi = \phi c_s^2 \tau_R$, instability requires $\phi > \gamma c^4/c_s^4$ for parallel or perpendicular propagation.  

In the comoving frame, the viscous modes are stable if $\Delta P = 0$ in the initial conditions.  All parallel and perpendicular modes are either neutral or damped. The stability for viscous modes in a non-comoving frame will be discussed in \ref{sec:noncomoving_frame}.

\subsection{Firehose Instability}

In non-relativistic Braginskii theory, the inclusion of anisotropic conduction and viscosity opens up avenues for qualitatively new types of linear instabilities.   For example, in the presence of anisotropic conduction, there are buoyancy instabilities driven by temperature gradients rather than entropy gradients  (\citealt{Balbus2000, Quataert2008}). In the presence of anisotropic viscosity, the character of the magnetorotational instability can change, with viscous transport of angular momentum driving an instability even when magnetic tension is negligible (\citealt{Quataert2002,Balbus2004}).  And, finally, a background pressure anisotropy can itself be subject to a host of instabilities, at least one of which (the firehose instability) is well captured by the EMHD-type models described here (e.g., \citealt{Schekochihin2009}). We expect that much of this rich physics will carry over to the relativistic theory described in this paper since it reduces to Braginskii theory in the non-relativistic limit.   As a simple concrete example of this, we derive here the relativistic instability criterion for the firehose instability.

Pressure anisotropy modifies the propagation speed of Alfv\'en waves and, if $\Delta P$ is large enough, transforms them into unstable, non-propagating modes.  Firehose instability has not appeared until now because our equilibrium has $\Delta P = 0$. But we can recover the firehose instability by setting $\Delta P \ne 0$ in the background state and taking $\tau_R \rightarrow \infty$ so that the pressure anisotropy cannot decay and the initial state is an equilibrium.  

The resulting dispersion relation for parallel propagation is:
\begin{equation}
\omega^2 = k^2 \frac{b^2 + \Delta P}{b^2 + \rho_0 + \gamma u_0 + \frac{1}{3}\Delta P}.
\end{equation} 
For $\Delta P = 0$ we recover the Alfv\'en waves.  For $\Delta P > 0$ the propagation speed increases and becomes superluminal if $\Delta P > \frac{3}{2} (\rho_0 + \gamma u_0)$.  However, this would require $\Delta P/P  > 3 \gamma/(2 (\gamma - 1))$.  For $\Delta P < 0$ we recover the relativistic firehose instability criterion:
\begin{equation}
b^2 + \Delta P < 0 \,\, \Rightarrow \, \, {\rm INSTABILITY}.
\end{equation}
This criterion is consistent with the relativistic kinetic criterion of \cite{Lerche1966}.

\subsection{Instability for Large $q/u$}

\cite{Hiscock1988} (hereafter HL88) have shown that the Israel-Stewart theory loses stability and hyperbolicity above a critical value of $q/u \simeq 0.08898$ in the ultrarelativistic ($u \gg \rho$) limit.  What does this imply for the EMHD model?

Consider an initial state with a background heat flux $q = q_0$.  We will perturb around this state to test stability.  But to do this, we are obliged to take the limit $\tau_R \rightarrow \infty$ so that the initial state with $q = q_0$ becomes an equilibrium.  It matters how this limit is taken: if we assume $\chi \propto \tau_R$, terms proportional to both $\tau_R$ and $\chi$ must be retained.
Linearization around this equilibrium for modes with $k \parallel B$ yields a fourth-order dispersion relation for the coupled sound/entropy/conduction excitations.

The stability of the $q = q_0$ equilibrium depends on $\gamma$, $\phi,q, \rho,$ and $u$ in a complicated way.  To compare with HL88 we consider the ultrarelativistic limit with $\gamma = 4/3$ and $u \gg \rho$; HL88 also make a choice for the coefficient $b_1$ that is equivalent to setting $\phi = 12/5$.  This simplifies the analysis, and one can show that the discriminant of the dispersion relation (which in this limit has real coefficients) changes sign at $q_0/u = 0.08898$, as in HL88. Indeed, one can show that HL88's $\lambda = 1$ Israel-Stewart model is identical to ours if we restrict attention to motion along the field.

We can generalize HL88's analysis by allowing $\phi$ to be different from $12/5$.  The critical $q_0/u = 2^{1/2}/3$ as $\phi  \rightarrow  0$.  Between $\phi = 0$ and $\phi = 3$ the critical $q_0/u$ is slightly above $q_0/u = (3 - \phi)/10$, still in the ultrarelativistic limit.

Is the instability a consequence of the second-order terms in the evolution equation for $q$?  The higher order terms enter the linear theory only when $q_0 \ne 0$, so they have played no role until now.  Turning these terms off and repeating the stability analysis, one finds the discriminant of the dispersion relation is positive definite. The instability is therefore seated in the second order terms. This is relevant for numerical implementations of the EMHD model, since it suggests that manipulating the higher order terms may improve stability, albeit at the cost of violating the second law.

\subsection{Stability in a non-comoving frame} \label{sec:noncomoving_frame}

It is an interesting feature of the Israel-Stewart theory that in some cases perturbations comoving with the fluid appear stable, but even a small relative velocity between the frame in which the perturbation is defined and the fluid, makes the perturbation unstable (\citealt{Hiscock1985}).  Consider a background velocity $u^\mu = (\Gamma,u^x,0,0)$, with $\Gamma = \sqrt{1+u_x^2}$ the Lorentz factor, a background magnetic field $b_\mu=(-b_x u^x/\Gamma ,b_x,0,0)$, and perturbations $(\delta \rho,\delta u_x,\delta \Delta P)$ of the density, longitudinal velocity and pressure anisotropy. From the linearly perturbed equations of mass conservation, evolution of the pressure anisotropy, and $\nabla_\mu \delta T^\mu_x=0$, we get (in the absence of heat conduction)
\begin{widetext}
\beqn
\Gamma \partial_t(\delta \rho) &=& \rho_0 u_x \partial_t(\delta u_x), \\
3\rho_0 h_0 (1+2u_x^2)\partial_t(\delta u_x) &=& 2 u_x \Gamma^2 \partial_t (\delta \Delta P)
- 3 u_x \Gamma^2 \partial_t (\delta \rho),\\
2\rho_0 \nu b_2 \Gamma^2 \partial_t (\delta \Delta P) &=& - \Gamma \delta \Delta P + 2\rho_0 \nu u_x \partial_t \delta u_x.
\eeqn
\end{widetext}
Substituting for $\partial_t \delta u_x$ and $\partial_t \delta \rho$, we get the evolution equation for $\delta \Delta P$,
\beq
\partial_t (\delta \Delta P) = - \frac{3(2\rho_0 (h_0-1) u_x^2 + \rho_0 h_0)}{3 b_2(\rho_0 h_0 \Gamma^2 + \rho_0 (h_0-1)u_x^2) - 2 u_x^2}\frac{\delta \Delta P}{2\rho_0 \nu \Gamma}.
\eeq
In this case, the stability condition is 
\beq
b_2> \frac{2 u_x^2}{3(\rho_0 h_0 \Gamma^2 +\rho_0 (h_0-1) u_x^2)}.
\eeq

Note that $b_2$ is the coefficient of the second order contribution to the entropy current (\ref{eq:entropyExpansion}) due to shear stress and is equal to $\tau_R/(2 \rho_0 \nu)$.

\subsection{Causality}
For small $\tau_R$, i.e., for small $b_1$ and $b_2$, the model is no longer causal. To demonstrate this, we consider the principal part of the evolution equations at a point at which $u_\mu=(-1,0,0,0)$, and $b_\mu=(0,b_x,0,0)$, and compute the characteristic speeds of the system in the tetrad frame and along the direction $\overrightarrow{e}_{(x)}$. Writing the linearized system of equations as 
\beq
A^t \partial_t (\delta U) + A^i \partial_i (\delta U) + B \delta U
\eeq
for perturbations $\delta U$, with matrices $A^\mu$, $B$ depending on the unperturbed variables $U_0$, then the characteristic speeds $v$ along direction $i$ are solutions of the equation
\beq
det(v*A^t - A^i)=0.
\eeq
In the absence of heat conduction, the relevant parts of the equations for $(\rho,u,u_x,\Delta P)$ are then
\beqn
\partial_t \rho &=& - \rho \partial_x u_x +\ ...\\
\partial_t u &=& - (u+P) \partial_x u_x + ...\\
\rho h \partial_t u_x &=& -\partial_x P + \frac{2}{3} \partial_x \Delta P +\ ... \\
\partial_t \Delta P &=& \frac{1}{b_2} \partial_x u_x +\ ...
\eeqn
(the transverse components of the magnetic field and velocity still follow the standard dispersion relation for Alfven waves).
The characteristic speeds of the system then depend on the equation of state $P(\rho,u)$. For a simple gamma-law fluid, $P=(\gamma-1)u$, we find the speeds $(0,0,\pm\sqrt{(2\rho \nu+3(\gamma-1)\gamma u b_2)/(3\rho h b_2)})$. For small values of $b_2$, two of the speeds are thus greater than $c=1$. To avoid superluminal speeds, we need
\beq
b_2 > \frac{2}{3(\rho+\gamma u (2-\gamma))}.
\eeq
Rewriting in terms of the relaxation time scale $\tau_R$, the condition for causality is
\beq
\tau_R > \frac{4 \rho \nu}{3(\rho+\gamma u (2-\gamma))}.
\eeq
This is identical to what we would have inferred from equation (\ref{eq:dispnul}) by imposing $\omega/k < c$.  Regardless of our choice of frame, the theory is thus problematic for small $b_2$, i.e. small $\tau_R$ or large $\nu/\tau_R$. These results are once more reminiscent of what \cite{Hiscock1983} found for an isotropic viscosity. In that case, they derived the stability condition $b_2>2/(3\rho h)$. \cite{Hiscock1983,Hiscock1988b} also demonstrated that the stability of the theory implied that the system of evolution equations was hyperbolic and causal, and that if the system of equations was either not causal or not hyperbolic, it was unstable.

\section{Nonlinear Theory}

In this section we give a brief but incomplete description of the behavior of the model in shocks.

Ideal hydrodynamics allows shocks and contact (entropy) discontinuities.  The boundary conditions at the discontinuity are determined by continuity conditions on the fluxes of mass, momenta, and energy.

Once viscosity is included in a non-relativistic fluid model the discontinuity is replaced by a sharp but continuous transition of width $\Delta x$ such that the diffusion timescale $\Delta x^2/\nu$ is comparable to the transit time through the shock $\Delta x/v$, so $\Delta x \sim \nu/v$.  If $\nu \sim \lambda_{mfp} c_s$ and $v \sim c_s$ then $\Delta x \sim \lambda_{mfp}$.  Put differently, viscosity can propagate information upstream at characteristic speed $\sim \nu/\Delta x$, which is supersonic if the structure is narrow enough.  Conduction produces an upstream precursor but does not remove the discontinuity \citep{Mihalas1984}.

In Maxwell-Cattaneo type theories there is a new characteristic speed, $\propto (\nu/\tau_R)^{1/2}$; if $\nu \sim c_s^2\tau_R$ then this speed is comparable to $c_s$.  If the upstream velocity exceeds this, the shock is still discontinuous.  In Israel-Stewart theory, this was studied by \cite{Olson1990}. See \cite{Bouras2010} for a numerical study of shocks in this theory and a comparison to shock solutions obtained using kinetic theory.

Linear analysis showed (\S\ref{sec:linear}) that our EMHD model contains a speed $\sim (\chi/\tau_R)^{1/2} \sim \phi^{1/2} c_s$ associated with thermal conduction and another speed $\sim \psi^{1/2} c_s$ associated with viscosity.  One would then expect the model to have a discontinuity over a sufficiently strong shock.  How, then, do $\Delta P$ and $q$ change across the discontinuity?  

The variation of $\Delta P$ and $q$ depend on the shock substructure, essentially because there are no continuity conditions, such as those present in ideal MHD, that can be used to navigate across the shock.  To clarify this point, consider the conduction model governed by equation (\ref{eq:simpqevol}) in the frame of a strong (so that $v$ upstream is greater than $v_q$) steady shock at $x = 0$, with $u^x > 0$.  The dominant terms near the shock are
\begin{equation}
u^x \partial_x Q = \frac{1}{\tau_R} \left(\frac{\tau_R}{\rho\chi P}\right)^{1/2} \left(-\rho \chi \hat{b}^x (\partial_x \Theta + \Theta u^x \del_x u_x)\right)
\end{equation}
because both $\partial_x T$ and $\partial_x u_x$ are large inside the shock.  This equation has the form
\begin{equation}
\partial_x Q = F(\rho,\Theta,u^x) \delta(x)
\label{eq:Qshock}
\end{equation}
where $\delta$ arises from $\partial_x$ acting on the discontinuities.  If $F$ were continuous through the shock we could integrate this equation across the jump and find a definite solution for $Q$ at $x > 0$.  But $F$ is changing discontinuously through the shock and hence the postshock $Q$ depends on the substructure of the shock.

EMHD shocks contain substructure on three scales.

(1) On small scales the width of the shock is set by bulk viscosity.  

Because bulk viscosity is not included in the model, this smallest scale is unresolved and the evolution of $q$ and $\Delta P$ across the shock is undetermined (but the change in, e.g., $Q$ across the shock $\equiv \Delta Q$ is limited, since if $F$ in equation (\ref{eq:Qshock}) is monotonic then $\Delta Q$ must lie between $F$ in the downstream state and $F$ in the upstream state).  In a numerical evolution it will be set by the numerical closure. 

(2) On intermediate scales the pressure anisotropy and heat flux relaxation produce a tail of width $\sim u^x \tau_R$.

The separation of scales (1) and (2) is an artifact of the model since if $\tau_R$ is small inside the shock, (1) and (2) can be comparable. 

(3) Far from the shock $\Delta P$ and $q \rightarrow 0$ and the shock obeys the ideal MHD jump conditions.  

The latter point implies that unless one is interested in shock substructure (and it would seem that EMHD is not the ideal model for collisionless shocks), the EMHD model may provide an adequate description of flow more than a few mean free paths from the shock.

\section{Discussion and Conclusion}

In this paper, we have derived an extended MHD model for a relativistic { weakly collisional} plasma  incorporating the effects of anisotropic conduction and viscosity. We are motivated by applications to low accretion rate black holes, but the model can also be applied in other contexts such as neutron star magnetospheres. This is the first in a series of publications. While this paper deals with the derivation, linear theory and stability analysis of our model, a later publication will discuss a flexible new code which we have written to solve this model numerically, as well as possible future extensions of this model. As tests for the code, and to get insights into the model, we derive various analytic/semi-analytic solutions to this model. The code, as well as the various tests will be described in \cite{Chandra2015}. Finally, we shall apply this model and the code to study the dynamics and structure of an accretion disk around a slowly accreting Kerr black hole.

\subsection{Summary of the Governing Equations}

 The complete model is given by the usual continuity equation (\ref{eq:continuity}), energy-momentum equations (\ref{eq:divT}) and induction equation (\ref{eq:induction}), supplemented by a heat conduction model (\ref{eq:qmodel}) and shear viscosity model (\ref{eq:taumodel}), together with evolution equations for the magnitude of the heat flux $q$  (\ref{eq:qevol}) and the pressure anisotropy $\Delta P$ (\ref{eq:dPevol}). 

\subsection{Summary of the Formalism}

We deduced the form of the heat flux (\ref{eq:qmodel}) and the viscous stress (\ref{eq:taumodel}) by examining the symmetries of the system in the presence of a magnetic field and by appealing to the small values of the gyroradius and the gyroperiod in astrophysical plasmas. The evolution equations (\ref{eq:qevol}) and (\ref{eq:dPevol}) are then derived using the Israel-Stewart theory of dissipative hydrodynamics. We have expanded the entropy current up to second order in deviations from equilibrium and then enforced its divergence to be positive. From this simple principle, we get equations for the  heat flux and the pressure anisotropy in a magnetized plasma. The derivation naturally shows that, as in non-relativistic plasmas, 1) the heat flux is driven by temperature gradients projected along the field lines and that 2) the pressure anisotropy is driven by a background shear projected along the field lines. Since our model is based on the Israel-Stewart theory, it conditionally satisfies all the requirements of dissipative theories in general relativity: hyperbolicity, causality and stability. For example, in the ultrarelativistic limit ($u \gg \rho$), the model is hyperbolic as long as $q/u < 0.089$.

In the limit where the relaxation time scale $\tau_R \rightarrow 0$, our model reduces to that of \cite{Braginskii1965}. The hyperbolic equations for parallel heat flux and pressure anisotropy relax to forms (\ref{eq:q0def}) and (\ref{eq:dP0def}) respectively, which are covariant generalizations of the Braginskii closure.  The pressure anisotropy in Braginskii theory arises due to a balance between the conservation of adiabatic invariants and pitch-angle scattering. In the appendix, we show using relativistic kinetic theory that this is true of our model as well. { In the non-relativistic limit, the inclusion of anisotropic conduction leads to many new types of instabilities, such as the magnetothermal instability (\citealt{Balbus2000}) and the heat flux driven buoyancy instability (\citealt{Quataert2008}), both of which are also modified by the inclusion of anisotropic viscosity (\citealt{Kunz2011})}.  Our relativistic model should also display all the instabilities that arise due to inclusion of  anisotropic dissipative effects. As an example, we showed that our model reproduces the correct threshold for the firehose instability. 

We relate the transport coefficients in our model ($\chi, \nu$) to the effective scattering time scale $\tau_R$. A collisionless plasma is subject to a number of kinetic instabilities (mirror, firehose, ion cyclotron, electron whistler) that effectively regulate the pressure anisotropy and the heat flux. Our closure prescription incorporates the isotropizing effects of these instabilities through the modification of $\tau_R$, thus providing a convenient way to incorporate the various kinetic effects. The exact way in which $\tau_R$ should be modified will ultimately be answered by first principle particle-in-cell calculations, which are well under way (\citealt{Kunz2014,Riquelme2015}).

{ \subsection{Two-Fluid Effects and Observational Signatures} }

The model is a one-fluid model of a plasma consisting of multiple species, here electrons and ions. The one-fluid matter stress tensor $T^{\mu \nu}_{\mathrm{matter}}$ in (\ref{eq:fullSEtens}) has been obtained by summing up stress tensors of each individual species in (\ref{eq:stressenergy}). Therefore, the rest mass density $\rho$, the internal energy $u$, the pressure $P$, the heat flux $q^\mu$ and the shear stress $\Pi^{\mu \nu}$ in (\ref{eq:fullSEtens}) are all a sum of the respective quantities of both electrons and ions (using $u^\mu_e \approx u^\mu_i \equiv u^\mu$), i.e. $\rho = \rho_e + \rho_i$, $u = u_e + u_i$, $P = P_e + P_i$, $q^\mu = q^\mu_e + q^\mu_i$ and $\Pi^{\mu \nu} = \Pi^{\mu \nu}_e + \Pi^{\mu \nu}_i$. The temperature  $\Theta$ in the model is a combination of both electron $\Theta_e$ and ion temperatures $\Theta_i$, i.e., $\Theta = (m_i \Theta_i + m_e \Theta_e)/(m_i + m_e)$ (since $P = P_e + P_i$), where $m_e$ and $m_i$ are the electron and ion masses, $\Theta_e \equiv k T_e/(m_ec^2)$, and $\Theta_i \equiv k T_i/(m_ic^2)$. Introduction of a single heat flux $q^\mu$ and a single shear stress $\Pi^{\mu \nu}$ for a system with multiple species is inconsistent, unless we make further approximations.

The ratio of the relaxed electron to ion heat flux is $q_{e0}/q_{i0} \sim (P_e/P_i)^2 (\tau_e/\tau_i) (m_i/m_e) (L_i/L_e)$, where $L_e$ and $L_i$ are the characteristic scales for the variation of the electron and ion temperatures and $\tau_e$ and $\tau_i$ are the electron and ion relaxation time scales respectively. Assuming $L_e \sim L_i$ and $\tau_e \sim \tau_i$, our one-fluid model is a reasonable description of the overall dynamics when $P_e/P_i \ll \sqrt{m_e/m_i}$. In this limit, the contribution to the total heat flux is dominated by the ion heat flux $q^\mu \approx q^\mu_i$, and the single temperature $\Theta$ in our one-fluid model is the ion temperature $\Theta \approx \Theta_i$. Similarly, the ratio of the relaxed electron to ion pressure anisotropy is $\Delta P_{e0}/\Delta P_{i0} \sim (P_e/P_i) (\tau_e/\tau_i) \ll 1$, when $P_e/P_i \ll 1$ (again assuming $\tau_e \sim \tau_i$). In this limit, the total pressure anisotropy is dominated by the ions and thus we have for the total shear stress $\Pi^{\mu \nu} \approx \Pi_i^{\mu \nu}$. Note that in the collisional (Braginskii) regime, $P_e \simeq P_i$, and $\tau_e/\tau_i \sim \sqrt{m_e/m_i}$, where $\tau_e$ and $\tau_i$ are the Coulomb scattering time scales for electrons and ions. The dominant heat flux is now due to electrons $q_{e0}/q_{i0} \sim \sqrt{m_i/m_e} \gg 1$, while the dominant pressure anisotropy is still due to ions $\Delta P_{e0} /\Delta P_{i0} \sim \sqrt{m_e/m_i} \ll 1$. Since in this collisional limit, $T_e \simeq T_i$, evolving a single temperature $\Theta \approx 2 \Theta_i$, as we do in this paper is appropriate.


Generalizing this one-fluid model to a two-fluid model that accounts for both electron and ion dissipation can be done in a number of ways. One approach is to use separate conservation equations for both the electrons and ions, into which our model can be incorporated in a straightforward manner. This however includes kinetic length scales in the system, which cannot be resolved in a global accretion disk simulation. 
Another approach is to work within the MHD ordering, which only requires one additional variable, the electron temperature.  \cite{Ressler2015} have developed such a model by using a separate electron entropy equation into which they incorporate electron conduction using a reduced version of the anisotropic conduction equation that we have derived here. The emission from the plasma is dominated by the electrons and hence the electron thermodynamical quantities are crucial to predict observables such as spectra, images and light curves.


The non-ideal effects modeled here may have a number of implications for the dynamics and observational properties of slowly accreting black holes: 1) conduction can redistribute energy spatially, changing where the emission comes from and potentially changing the dynamics by increasing the pressure at large latitude/radii and 2) viscosity can provide an additional source of transport and heating (over and above the Maxwell and Reynolds stress), potentially modifying $\dot{M}$ and the thermodynamics of the plasma -- the latter being relevant for the emission.

\subsection{Connection to Other Models}

The model we have presented is the simplest in a hierarchy of possible models  incorporating non-ideal effects in magnetized plasmas. In particular, it only includes a single parallel heat flux $q^\mu = q \,\hat{b}^\mu$ and the parallel and perpendicular pressures are controlled by a single variable, the pressure anisotropy $\Delta P$. The derivation was based on  thermodynamic principles. On the other hand, models of collisionless plasmas are usually derived by taking velocity space moments of the Vlasov equation. Early examples of relativistic models derived in this way are by \cite{Gedalin1991}, \cite{Tsikarishvili1992} and \cite{Tsikarishvili1994}. They have independent variables for the parallel ($P_\parallel$) and perpendicular ($P_\perp$) pressures. However, their stress tensors lack a heat flux and their models reduce in the non-relativistic limit to the well-known CGL closure (\citealt{Chew1956}). A more recent model by \cite{TenBarge2008} has evolution equations for $P_\parallel$ and $P_\perp$ and these couple individually to separate heat fluxes $q_\parallel \sim \hat{b}^\mu \nabla_\mu P_\parallel$ and $q_\perp \sim \hat{b}^\mu \nabla_\mu P_\perp$. In the non-relativistic regime,  when the mean free path is small compared to the system size, such a model recovers the Braginskii limit (\citealt{Snyder1997}). Thus, we expect our model to be formally applicable only in this regime. Also, since we have not derived our model using kinetic theory, we do not expect that it will reproduce all of the  kinetic linear modes; for example, our model lacks the mirror instability. However, a nice feature of our model is that it includes collisions whereas the previous relativistic models for collisionless plasmas do not. To include collisions in the moment formalism that the previous models used, one has to include a collision operator in the Vlasov equation and take its moments. Our model naturally includes collisions because it is an expansion up to second order around thermal equilibrium. The relaxation time scale $\tau_R$ can be interpreted as the relaxation time scale in the BGK collision operator. Its presence is convenient since it makes it easy to incorporate subgrid models of the saturation of kinetic plasma instabilities as an effective collisionality. 

Our model captures the leading order effects of heat transport and pressure anisotropy while still being relatively simple. More sophisticated models have been derived from Israel-Stewart theory in the presence of a magnetic field (\citealt{Huang2010}), albeit in a different context, for use in strange quark stars with strong magnetic fields. The Israel-Stewart theory itself is a specific instance of a class of theories derived from Extended Thermodynamics, which are based on an entropy principle, such as the one we have used in this paper. See \cite{Jou1988} for a review. While this formalism gives us the evolution equations for the dissipative fluxes, one needs to eventually resort to kinetic theory in order to compute the transport coefficients. To perform this computation, \cite{Israel1979} have used an approach similar to that of \cite{Grad1949}. They expand the distribution function in momentum space polynomials around an equilibrium, i.e., $f = f_0(1 + a_\mu p^\mu + b_{\mu \nu}p^\mu p^\nu)$, where $f_0$ is the equilibrium distribution function. This ansatz for the distribution function is then used along with the second moment of Boltzmann equation to compute the transport coefficients, which come out in terms of moments of $f_0$. Note that the second moment of the Boltzmann equation is one moment higher than the divergence of the stress-tensor and, it is at this level of the moment hierarchy where the collision operator makes a non-zero contribution.

\acknowledgments
MC is supported by the Illinois Distinguished Fellowship from the University of Illinois.
This work was supported by NSF grant AST 13-33612 and NASA grant NNX10AD03G, and a Romano Professorial Scholarship to CFG. 
Support for this work was provided by NASA through Einstein Postdoctoral Fellowship grant PF4-150122 to FF awarded by the Chandra X-ray Center, which is operated by the Smithsonian Astrophysical Observatory for NASA under contract NAS8-03060. 
EQ is supported in part by a Simons Investigator Award from the Simons Foundation and the David and Lucile Packard Foundation.
We thank B. Ryan and S. Shapiro for discussions as well as all the members of the horizon collaboration, {\tt horizon.astro.illinois.edu}, for their advice and encouragement. Some of the linear theory calculations were perfomed using the {\tt sagemath} software on {\tt sagemathcloud.com}.
\begin{appendix}
\section{Adiabatic Invariants and Pressure Anisotropy}
\label{sec:adiabatic}

Equation \ref{eq:dP0def} in the main text for $\Delta P_{0}$ is the natural general relativistic generalization of the well-known \citet{Braginskii1965} relation between pressure anisotropy, viscosity, and shear stress in a non-relativistic plasma.  In the non-relativistic limit, this relationship also has a simple interpretation in terms of a balance between (1) the rate of generation of pressure anisotropy by adiabatic invariance of the magnetic moment in a slowly varying magnetic field and (2) the isotropization in velocity space by pitch-angle angle scattering at a rate $\nu_s$.   We briefly review this non-relativistic result and then discuss its generalization to relativistic kinetic theory of a magnetized plasma.   This provides a useful physical interpretation of the thermodynamic derivation of the equilibrium pressure anisotropy (eq. \ref{eq:dP0def}) in the main text.  We take $k = m = c = 1$ throughout this Appendix.

\subsection{Non-relativistic Kinetic Theory}

The viscous stress tensor in a magnetized non-relativistic collisional plasma in which the cyclotron frequency is much larger than the collision frequency is given by \citep{Braginskii1965}
\begin{equation}
\mathbf{\Pi} = -3\rho\nu \left[\bhat\bhat:\grad \boldsymbol{v}
-\frac{\grad\cdot\boldsymbol{v}}{3}\right]
\left[\bhat\bhat - \frac{\mathbf{I}}{3}\right],
\label{eq:viscous-tensor}
\end{equation}
Equation \ref{eq:viscous-tensor} can also be written as
\begin{equation}
\mathbf{\Pi} = - \Delta P
\left[\bhat\bhat - \frac{\mathbf{I}}{3}\right],
\label{eq:viscous-tensor2}
\end{equation}
where the pressure anisotropy in the non-relativistic limit  is given by
\begin{equation}
\Delta P = 3\rho\nu \left[\bhat\bhat:\grad \boldsymbol{v}
-\frac{\grad\cdot\boldsymbol{v}}{3}\right].
\label{eq:dP}
\end{equation}
$\Delta P$ in equations \ref{eq:viscous-tensor2} and \ref{eq:dP} is defined as $\Delta P = P_\perp - P_\parallel$, with directions defined by the local magnetic field in the plasma.   
Note that since $\nu \sim c_s^2/\omega$ where { $\omega$ is the pitch angle scattering rate} and $c_s$ is the sound speed, equation \ref{eq:dP} is equivalent to 
\begin{equation}
\omega \, \frac{\Delta T}{T} \sim \frac{d}{dt} \ln\left[\frac{B^3}{\rho^2}\right]
\label{eq:dP-ad}
\end{equation}
where we have used the induction equation and mass conservation to rewrite the right-hand-side of equation \ref{eq:dP-ad}.   The left hand side of equation \ref{eq:dP-ad} is the rate at which scattering at rate $\omega$ decreases the pressure anisotropy.   The right hand side of equation \ref{eq:dP-ad} is the rate at which adiabatic invariance of $T_\perp/B$ (magnetic moment) and $T_\parallel B^2/\rho^2$ (the `bounce' invariant) change the pressure anisotropy.    The assumption of collisional theory is that these two effects approximately balance each other. 

\subsection{Relativistic Kinetic Theory}

The thermodynamic derivation in \S \ref{sec:evolution} (in particular eq. \ref{eq:dP0def}) shows that results very similar to equations \ref{eq:dP} and \ref{eq:dP-ad} relate pressure anisotropy and viscosity in GR.   To understand the microscopic origin of these results, it is instructive to consider the relativistically correct first and second adiabatic invariants for an individual particle \citep{Sturrock1994}:
\begin{equation}
\frac{p_\perp^2}{b} = {\rm constant} \hspace{0.5in} \frac{p_\parallel b}{\rho} = {\rm constant}
\label{eq:rel-adiabatic}
\end{equation}
where $p_\perp$ and $p_\parallel$ are the particle momenta (not pressure!) along and perpendicular to the magnetic field direction defined in the fluid frame.  What does this imply for the relation between $\Delta P$ and the variation of $b$ and $\rho$?

\def\jp{{j_\perp}}
\def\jl{{j_\parallel}}

Start by changing momentum space coordinates to the adiabatic invariants
\begin{equation}
\jp \equiv \frac{p_\perp^2}{m} \hspace{0.5in}
\jl \equiv p_\parallel \left(\frac{m}{r}\right)
\end{equation}
where $r \equiv \rho/\rho_0$ and $m \equiv b/b_0$.  The initial ($r = m = 1$) Maxwellian distribution function is 
\begin{equation}
\frac{d n}{d\jp d\jl} = \frac{n}{4\Theta K_2(1/\Theta)} e^{-\Gamma/\Theta}
\end{equation}
where 
\begin{equation}
\Gamma = (1 + \jp + \jl^2)^{1/2}
\end{equation}
and $K_2$ is the modified Bessel function of the second kind. This distribution is invariant under slow (adiabatic) changes in $b$ and $\rho$.

Next, directly evaluate components of the pressure tensor using the kinetic theory definition (\ref{eq:ptensdef}) in the limit that $r$ and $m$ are close to $1$:
\begin{equation}
\Delta P = P_\perp - P_\parallel = \int \, d\jp d\jl \, \frac{d n}{d\jp d\jl} \, \left( \frac{1}{2}\frac{p_\perp^2}{p^t} - \frac{p_\parallel^2}{p^t}\right),
\end{equation}
where the factor of $\frac{1}{2}$ arises because $p_\perp^2 = p_x^2 + p_y^2$ if $b$ is aligned with $z$; here $p^t = \Gamma$ in the fluid frame.  Notice that $p_\perp$ and $p_\parallel$ depend on $r$ and $m$ but the distribution function does not.  Then
\begin{equation}
\left(\frac{d\Delta P}{d\tau}\right)_{ad} = 
\frac{d\Delta P}{dm}\frac{d m}{d\tau} +
\frac{d\Delta P}{dr}\frac{d r}{d\tau}
\end{equation}
is the change in $\Delta P$ due to adiabatic deformation of the distribution function.  If the anisotropy is small the derivative can be evaluated at $b = m = 1$.

The integrals needed for $d\Delta P/dm$ and $d\Delta P/dr$ can be evaluated analytically in the relativistic and non-relativistic limits.

In the non-relativistic limit $\Gamma \rightarrow 1 + \frac{1}{2}\jp + \frac{1}{2}\jl^2$ and $dn/d\jp d\jl \propto \Theta^{-3/2} \exp(-(\jp + \jl^2))/(2\Theta)$.  A few strengthening integrals later, one finds
\begin{equation}
\frac{1}{P}\left(\frac{d\Delta P}{d\tau}\right)_{ad} = 
3 \frac{d m}{d\tau} - 
2 \frac{d r}{d\tau} 
\end{equation}
where $P = n\Theta$.  Since the derivatives are evaluated at $m = r = 1$, $d m/d\tau = d\ln m/d\tau$, etc., and using the definition of $m,r$, find
\begin{equation}
\frac{1}{P}\left(\frac{d\Delta P}{d\tau}\right)_{ad} = 
\frac{d}{d\tau} \ln \left(\frac{b^3}{\rho^2}\right)
\end{equation}
as expected from the discussion in the preceding subsection.

In the ultrarelativistic limit $\Gamma \rightarrow (p_\perp^2 + p_\parallel^2)^{1/2} = (\jp + \jl^2)^{1/2}$ at $b = m = 1$, and $dn/d\jp d\jl \propto \Theta^{-3} \exp(-\Gamma/\Theta)$.  The integrals give
\begin{equation}
\frac{1}{P}\left(\frac{d\Delta P}{d\tau}\right)_{ad} = 
\frac{12}{5} \frac{d m}{d\tau} - 
\frac{8}{5} \frac{d r}{d\tau},
\end{equation}
or
\begin{equation}
\frac{1}{P}\left(\frac{d\Delta P}{d\tau}\right)_{ad} = 
\frac{4}{5}\frac{d}{d\tau} \ln \left(\frac{b^3}{\rho^2}\right).
\end{equation}
which differs by a factor of $4/5$ from the non-relativistic limit.

The rate of decay of the pressure anisotropy due to scattering is
\begin{equation}
\frac{1}{P}\left(\frac{d\Delta P}{d\tau}\right)_{scat} = -\omega \frac{\Delta P}{P},
\end{equation}
{ which defines the scattering rate $\omega$},
so the total change
\begin{equation}
\frac{1}{P}\left(\frac{d\Delta P}{d\tau}\right) = 
\frac{1}{P}\left(\frac{d\Delta P}{d\tau}\right)_{ad} +
\frac{1}{P}\left(\frac{d\Delta P}{d\tau}\right)_{scat}.
\end{equation}
In equilibrium scattering balances adiabatic forcing and, for the  ultrarelativistic limit, the result is
\begin{equation}
\omega \frac{\Delta P}{P} = 
\frac{4}{5}\frac{d}{d\tau} \ln \left(\frac{b^3}{\rho^2}\right).
\end{equation}
We can do just a little bit more by evaluating the right hand side using the induction equation and the continuity equation, together with $b_\mu u^\mu = 0$.  Along the way,
\begin{equation}
\frac{1}{2}\frac{d}{d\tau}\ln b^2 = - \nabla_\mu u^\mu + \bh^\mu \bh^\nu \nabla_\mu u_\nu
\end{equation}
and
\begin{equation}
\frac{d}{d\tau}\ln \rho = -\nabla_\mu u^\mu,
\end{equation}
so that
\begin{equation}
\frac{d}{d\tau} \ln \left(\frac{b^3}{\rho^2}\right) = 3 \left(\bh^\mu \bh^\nu \nabla_\mu u_\nu - \frac{1}{3} \nabla_\mu u^\mu\right).
\end{equation}
{
Then in equilibrium
\begin{equation}
\Delta P = 3 \rho\, [\frac{4}{5} \frac{\Theta}{\omega}] \,\left(\bh^\mu \bh^\nu \nabla_\mu u_\nu - \frac{1}{3} \nabla_\mu u^\mu\right). 
\end{equation}
This is consistent with the non-relativistic derivation in equation \ref{eq:dP}, and, if we set
\begin{equation}
\nu = \frac{4}{5} \frac{\Theta}{\omega} \propto \tau c_s^2
\end{equation}
}
with the relativistic thermodynamic derivation in \S \ref{sec:evolution} of the main text (in particular, equation \ref{eq:dP0def} and surrounding results).

\end{appendix}


\begin{thebibliography}{}

\bibitem[Anderson \& Kox(1977)]{Anderson1977} Anderson, J.~L., \& Kox, A.~J.\ 1977, Physica A Statistical Mechanics and its Applications, 89, 408 

\bibitem[Andersson \& Comer(2007)]{Andersson2007} Andersson, N., \& Comer, G.~L.\ 2007, Living Reviews in Relativity, 10, 1

\bibitem[Balbus \& Hawley(1991)]{Balbus.Hawley.1991} Balbus, S.~A., \& Hawley, J.~F.\ 1991, \apj, 376, 214

\bibitem[Balbus(2000)]{Balbus2000} Balbus, S.~A.\ 2000, ApJ, 534, 420

\bibitem[Balbus(2004)]{Balbus2004} Balbus, S.~A.\ 2004, ApJ, 616, 857

\bibitem[Bale et al.(2009)]{Bale2009} Bale, S.~D., Kasper, J.~C., Howes, G.~G., et al.\ 2009, Physical Review Letters, 103, 211101

\bibitem[Bouras et al.(2010)]{Bouras2010} Bouras, I., Molnar, E., Niemi, H., Xu, Z., El, A., Fochler, O., Greiner, C., \& Rischke, D.~H.\ 2010, Phys. Rev. C 82,
024910

\bibitem[Braginskii(1965)]{Braginskii1965} Braginskii, S.~I.\ 1965, Reviews of Plasma Physics, 1, 205

\bibitem[Cattaneo(1948)]{Cattaneo1948} Cattaneo, C.\ 1948, Atti Semin. Mat. Fis. Univ. Modena 3:83–101

\bibitem[Cercignani \& Kremer(2001)]{Cercignani2001} Cercignani, C., \& Kremer, G.~M.\ 2001, Physica A Statistical Mechanics and its Applications, 290, 192

\bibitem[Chandra et al.(2015)]{Chandra2015} Chandra, M., Foucart, F., \& Gammie, C.~F. (in prep)

\bibitem[Chew et al.(1956)]{Chew1956}  Chew, G.~F., Goldberger, M.~L., \& Low, F.~E.\ 1956 Proc. R. Soc. A, 236, 1204

\bibitem[Doeleman et al.(2008)]{Doeleman2008} Doeleman, S.~S., Weintroub, J., Rogers, A.~E.~E., et al.\ 2008, \nat, 455, 78

\bibitem[Doeleman et al.(2009)]{Doeleman2009} Doeleman, S., Agol, E., Backer, D., et al.\ 2009, astro2010: The Astronomy and Astrophysics Decadal
Survey, 2010, 68

\bibitem[De Villiers et al.(2003)]{DeVilliers2003} De Villiers, J.-P., Hawley, J.~F., \& Krolik, J.~H.\ 2003, \apj, 599, 1238

\bibitem[Eckart(1940)]{Eckart1940} Eckart, C.\ 1940 Phys. Rev. 58, 919

\bibitem[Eisenhauer et al.(2008)]{Eisenhauer2008} Eisenhauer, F., Perrin, G., Brandner, W., et al.\ 2008, \procspie, 7013, 70132A

\bibitem[El et al.(2010)]{El2010} El, A., Xu, Z., \& Greiner, C.\ 2010, Phys. Rev. C, 81, 041901

\bibitem[Gammie et al.(2003)]{gammie2003} Gammie, C.~F., McKinney, J.~C., \& T{\'o}th, G.\ 2003, \apj, 589, 444 

\bibitem[Garcia-Perciante et al.(2009)]{Garcia2009}  Garcia-Perciante, A.~L., Garcia-Colin, L.~S., Sandoval-Villalbazo, A.\ 2009, Gen. Rel. and Grav., 41, 7

\bibitem[Gedalin(1991)]{Gedalin1991} Gedalin, M.\ 1991, Phys. Fluids B 3, 1871

\bibitem[Goldreich \& Sridhar(1995)]{GS1995} Goldreich, P., \& Sridhar, S.\ 1995, \apj, 438, 763

\bibitem[Grad(1949)]{Grad1949} Grad, H.\ 1949, Comm. Pure Appl. Math., 2:331–407

\bibitem[Hellinger et al.(2006)]{Hellinger2006} Hellinger, P., Tr{\'a}vn{\'{\i}}{\v c}ek , P., Kasper, J.~C., \& Lazarus, A.~J.\ 2006, Geophys. Res. Lett., 33, L09101

\bibitem[Hellinger et al.(2015)]{Hellinger2015} Hellinger, P., Tr{\'a}vn{\'{\i}}{\v c}ek, P.~M., \& Tr{\'a}vn{\'{\i}}{\v c}ek 2015, Journal of Plasma Physics, 81, 3003

\bibitem[Hiscock \& Lindblom(1983)]{Hiscock1983} Hiscock, W.~A., \& Lindblom, L.\ 1983, Annals of Physics, 151, 466

\bibitem[Hiscock \& Lindblom(1985)]{Hiscock1985} Hiscock, W.~A., \& Lindblom, L.\ 1985, \prd, 31, 725

\bibitem[Hiscock \& Lindblom(1988)]{Hiscock1988} Hiscock, W.~A., \& Lindblom, L.\ 1988, Physics Letters A, 131, 509

\bibitem[Hiscock \& Lindblom(1988b)]{Hiscock1988b} Hiscock, W.~A., \& Lindblom, L.\ 1988, Contemporary Mathematics, 71, 181-220

\bibitem[Ho(2009)]{Ho2009} Ho, L.~C\ 2009, ApJ, 699, 626

\bibitem[Huang et al.(2010)]{Huang2010} Huang, X.~G., Huang, M., Rischke, D.~H., \& Sedrakian, A.\ 2010, Phys. Rev. D 81, 045015

\bibitem[Israel \& Stewart(1979)]{Israel1979} Israel, W., \& Stewart, J.~M.\ 1979, Annals of Physics, 118, 341

\bibitem[Johnson \& Quataert(2007)]{Johnson2007} Johnson, B.~M., \& Quataert, E.\ 2007, \apj, 660, 1273

\bibitem[Jou et al.(1988)]{Jou1988} Jou, D., Casas-Vazquez, J., \& Lebon, G.\ 1988, Rep. Prog. Phys., 51, 1105

\bibitem[Koide et al.(1999)]{Koide1999} Koide, S., Shibata, K., \& Kudoh, T.\ 1999, \apj, 522, 727

\bibitem[Kulsrud(2005)]{Kulsrud2005} Kulsrud, R.~M.\ 2005, Plasma physics for astrophysics / Russell M.~Kulsrud.~Princeton, N.J.~: Princeton University Press.

\bibitem[Kunz et al.(2014)]{Kunz2014} Kunz, M.~W., Schekochihin, A.~A., \& Stone, J.~M.\ 2014, Physical Review Letters, 112, 205003

\bibitem[Kunz(2011)]{Kunz2011} Kunz, M.\ 2011, MNRAS, 411, 1

\bibitem[Lerche(1966)]{Lerche1966} Lerche, I.\ 1966, ApJ, 145, 806

\bibitem[Landau \& Lifshitz(1987)]{Landau1987} Landau, L.~D., \& Lifshitz, E.~M., Fluid dynamics, Butterworth-Heinemann, 1987

\bibitem[Lopez-Monsalvo \& Andersson(2011)]{Lopez2011} Lopez-Monsalvo, C.~S., Andersson, N.\ 2011 Proc. R. Soc. A 2011 467 738-759

\bibitem[Mahadevan \& Quataert(1997)]{Mahadevan1997} Mahadevan, R., Quataert, E.\ 1997, \apj, 490, 605

\bibitem[McKinney \& Gammie(2004)]{McKinney2004} McKinney, J.~C., \& Gammie, C.~F.\ 2004, \apj, 611, 977

\bibitem[Mihalas \& Weibel Mihalas(1984)]{Mihalas1984} Mihalas, D., \& Weibel Mihalas, B.\ 1984, New York: Oxford University Press, 1984

\bibitem[Misner et al.(1973)]{Misner1973} Misner, C.~W., Thorne, K.~S., \& Wheeler, J.~A.\ 1973, San Francisco: W.H.~Freeman and Co., 1973

\bibitem[Moscibrodzka et al.(2014)]{moscibrodzka2014} Moscibrodzka, M., Falcke, H., Shiokawa, H., \& Gammie, C.~F.\ 2014, arXiv:1408.4743 

\bibitem[Olson \& Hiscock(1990)]{Olson1990} Olson, T.~S., \& Hiscock, W.~A.\ 1990, Annals of Physics, 204, 331

\bibitem[Pistinner \& Eichler(1998)]{Pistinner1998} Pistinner, S.~L., \& Eichler, D.\ 1998, \mnras, 301, 49

\bibitem[Psaltis et al.(2014)]{Psaltis2014} Psaltis, D., Ozel, F., Chan, C.-K., \& Marrone, D.~P.\ 2014, arXiv:1411.1454

\bibitem[Quataert(1998)]{Quataert1998} Quataert, E.\ 1998, \apj, 500, 978

\bibitem[Quataert et al.(2002)]{Quataert2002} Quataert, E., Dorland, W., Hammett, G.~W.\ 2002, 2002 ApJ 577 524

\bibitem[Quataert(2008)]{Quataert2008} Quataert, E.\ 2008, ApJ, 673, 758

\bibitem[Ressler et al.(2015)]{Ressler2015} Ressler, S.~M., Tchekhovskoy, A., Quataert, E., Chandra, M., Gammie, C.~F.~G, (under review)

\bibitem[Riquelme et al.(2015)]{Riquelme2015} Riquelme, M.~A., Quataert, E., \& Verscharen, D.\ 2015, \apj, 800, 27

\bibitem[Sharma et al.(2006)]{Sharma2006} Sharma, P., Hammett, G.~W., Quataert, E., \& Stone, J.~M.\ 2006, \apj, 637, 952

\bibitem[Sharma et al.(2007)]{Sharma2007} Sharma, P., Quataert, E., Hammett, G.~W., \& Stone, J.~M.\ 2007, \apj, 667, 714

\bibitem[Sironi \& Narayan(2015)]{Sironi2015} Sironi, L., \& Narayan, R.\ 2015, \apj, 800, 88S

\bibitem[Snyder et al.(1997)]{Snyder1997} Snyder, P.~B., Hammett, G.~W., \& Dorland, W.\ 1997 Phys. Plasmas 4, 3974

\bibitem[Stix(1992)]{Stix1992} Stix, T.~H.\ 1992, Waves in plasmas , by Stix, Thomas Howard.; Stix, Thomas Howard.~ New York : American Institute of Physics, c1992.

\bibitem[Sturrock(1994)]{Sturrock1994} Sturrock, P.~A.\ 1994, Plasma Physics, An Introduction to the Theory of Astrophysical, Geophysical and Laboratory Plasmas, ISBN 0521448107, Cambridge University Press, 1994

\bibitem[Schekochihin et al.(2009)]{Schekochihin2009} Schekochihin, A.~A., Cowley, S.~C., Dorland, W., Hammett, G.~W., Howes, G.~G., Quataert, E., \& Tatsuno, T.\  2009, ApJS, 182, 310

\bibitem[TenBarge et al.(2008)]{TenBarge2008} TenBarge, J.~M., Hazeltine, R.~D., \& Mahajan, S.~M.\ 2008, Phys. Plasmas 15, 062112


\bibitem[Tsikarishvili et al.(1992)]{Tsikarishvili1992} Tsikarishvili, E.~G., Lominadze, J.~G., Rogava, A.~D., \& Javakhishvili, J.~I.\ 1992, Phys. Rev. A 46,
1078

\bibitem[Tsikarishvili et al.(1994)]{Tsikarishvili1994} Tsikarishvili, E.~G., Lominadze, J.~G., \& Javakhishvili, J.~I.\ 1994, Phys. Plasmas 1, 150

\bibitem[Yan \& Lazarian(2004)]{Yan2004} Yan, H., \& Lazarian, A.\ 2004, \apj, 614, 757

\bibitem[Yuan \& Narayan(2014)]{Yuan2014} Yuan, F., \& Narayan, R.\ 2014, \araa, 52, 529



\end{thebibliography}

\end{document}